\documentclass{./aa}
\usepackage{graphicx}
\RequirePackage{natbib}
\bibpunct{(}{)}{;}{a}{}{,}

\newcommand{\fei}{\ion{Fe}{i}}
\newcommand{\bzavpl}{$\langle B_z \rangle=200\,{\rm G}$}
\newcommand{\bzavdec}{$\langle B_z \rangle=10\,{\rm G}$}

\begin{document}
\title{Stokes diagnostics of simulated solar magneto-convection }
\titlerunning{Stokes diagnostics}
\author{S. Shelyag \inst{1,2}
\and M. Sch\"{u}ssler \inst{2}
\and S.K. Solanki \inst{2}
\and A. V\"{o}gler \inst{2}}

\institute{{$^1$} Solar Physics and Space Plasma Research Centre,
Department of Applied Mathematics, University of Sheffield, Hicks
Building, Hounsfield Rd, S3 7RH, UK \\ 
{$^2$} Max-Planck-Institut f\"ur
Sonnensystemforschung, 37191 Katlenburg-Lindau, Germany}

\date{\today} \abstract{We present results of synthetic
spectro-polarimetric diagnostics of radiative MHD simulations of solar
surface convection with magnetic fields. Stokes profiles of
Zeeman-sensitive lines of neutral iron in the visible and infrared
spectral ranges emerging from the simulated atmosphere have been
calculated in order to study their relation to the relevant physical
quantities and compare with observational results. We have analyzed the
dependence of the Stokes-$I$ line strength and width as well as of the
Stokes-$V$ signal and asymmetries on the magnetic field strength.
Furthermore, we have evaluated the correspondence between the actual
velocities in the simulation with values determined from the Stokes-$I$
(Doppler shift of the centre of gravity) and Stokes-$V$ profiles
(zero-crossing shift).  We confirm that the line weakening in strong
magnetic fields results from a higher temperature (at equal optical
depth) in the magnetic flux concentrations.  We also  confirm that
considerable Stokes-$V$ asymmetries originate in the peripheral parts of
strong magnetic flux concentrations, where the line of sight cuts
through the magnetopause of the expanding flux concentration into the
surrounding convective donwflow.

\keywords{Sun: magnetic fields - Sun: photosphere}
}

\maketitle

\section{Introduction}

Measurements of the photospheric magnetic fields of the Sun are mostly
based on polarimetric observations of spectral lines \citep[for an
overview, see][]{solankirev,Solanki:etal:2006}. In particular, two line
pairs of neutral iron (\ion{Fe}{i}), one in the visible wavelength range
at 630.15~nm (effective Land{\'e} factor $g_{\mathrm eff}=1.66$) and
630.25~nm ($g=2.5$), and one in the infrared at 1564.85~nm ($g=3$) and
1565.28~nm ($g_{\mathrm eff}=1.53$) have often been used for
polarimetric studies since they are not blended by other lines, have
large Land{\'e} factors, and because the small wavelength separation of
each pair permits simultaneous observations with spectrographs
\citep[e.g.,][]{Stenflo:etal:1987b, Rabin:1992, solankietal1,
Lites:etal:1996, Martinez-Pillet:etal:1997, Sigwarth:etal:1999,
Stolpe:Kneer:2000, Socas-Navarro:Lites:2004}.  For magnetic field
diagnostics in the visible spectral range, the \ion{Fe}{i} lines at
630.15~nm and 630.25~nm have the advantage of being less temperature
sensitive than the other widely used line pair 524.7~nm and 525.0~nm
\citep[e.g.,][]{Stenflo:1973,Wiehr:1978,Stenflo:etal:1987,
Grossmann-Doerth:etal:1996}.  Compared to the line pair in the
visible, the infrared lines considered here are formed deeper in the
atmosphere. Infrared spectral lines with sufficiently large Land{\'e}
factors are well suited for the diagnostics of weak magnetic fields
since the Zeeman splitting grows proportionally to the square of the
wavelength while the Doppler width increases only linearly.

In this paper we carry spectro-polarimetric diagnostics of  3D
radiative MHD simulations obtained with the {\em MURaM} code
\citep{Voegler:etal:2003,voegler}, concentrating on the lines
\ion{Fe}{i} 630.25~nm in the visible and \ion{Fe}{i} 1564.85~nm in the
infrared.  While the analysis of \citet{Khomenko:etal:2005a,
Khomenko:etal:2005b} considered mixed-polarity regions with zero net
vertical flux through the computational box, our study is based upon two
simulations with non-vanishing net flux corresponding to horizontally
averaged magnetic field strengths of $200~\mathrm{G}$ and
$10~\mathrm{G}$, respectively. These simulations are considered to
represent small parts of active regions in early (plage) and late
stages, respectively, of their development. We calculate synthetic
profiles of the Stokes parameters for the visible and infrared \fei\
lines based upon the physical quantities in the simulated solar
photosphere, analyze their relation to the atmospheric structure, and
compare them with observational data.  In particular, we statistically
analyze the origin of asymmetries of Stokes V profiles originating from
different regions in the computational box and relate them to the
velocity and magnetic field profiles along the corresponding lines of
sight.

The paper is organized as follows. Sect.~2 gives a brief description of
the simulations underlying our analysis. The line synthesis on the basis
of the simulation data is explained in Sect.~3. The results are
presented and discussed in Sect.~4, and Sect.~5 summarizes our
conclusions.

\section{MHD simulations}

The MURaM code \citep{Voegler:2003,voegler} integrates the system of MHD
equations on a three-dimensional, equidistant cartesian grid. The code
is parallelized using a domain decomposition scheme. The upper boundary
of the computational domain is assumed to be closed. It is located in
the upper photosphere near the temperature minimum, where the density is
rather low, so that the influence of the closed boundary on the dynamics
of the granulation is negligible. The lower boundary is located in the
convectively unstable layers of the upper convection zone. An open
boundary is implemented in order to allow free motions of the fluid
through the boundary. The code includes non-grey (opacity binning)
radiative transport and an equation of state taking into account partial
ionization of the 11 most abundant chemical elements in the solar
photosphere.

The size of the computational domain for the simulations considered here
is $6000 \times 6000 \times 1400 \rm{~km}^3$ with a grid size of $288
\times 288 \times 100$ grid cells. The corresponding resolution is
$20.8\,$km in the horizontal directions and 14 km in the vertical. The
simulation starts with a plane-parallel atmosphere extending between 800
km below and 600 km above the average level of continuum optical depth
unity at $500 \rm{~nm}$.  After convection has developed and both the
outgoing energy flux and the total kinetic energy have reached
stationary values (with only short-term fluctuations on the granulation
time scale), a homogeneous vertical magnetic field of 200~G and 10~G,
respectively, has been imposed. After the decay of all transients caused
by the introduction of the magnetic field, we have continued the
simulations for about another hour solar time to be sure that the
results have become independent of the initial distribution of magnetic
flux.

The snapshot from the 200~G run (representing a small part of a plage
region) is taken about 160 minutes after the start of simulation and 95
minutes after the vertical magnetic field was introduced. The snapshot
from the 10~G run (late decay phase of an active region, or base of a
coronal hole) is taken about 165 minutes after the start of the
simulation and 45 minutes after the vertical magnetic field was
introduced.  Maps of the vertical components of magnetic field and flow
velocity on the surface $\tau_{500}=1$ for the two simulation snapshots
are shown in Fig.~\ref{simpic}. They illustrate the concentration of the
magnetic flux in the intergranular downflow lanes and the formations of
flux sheets (mainly in the 200~G snapshot) and tube-like magnetic
structures.

\section{Line synthesis}

We have calculated Stokes profiles for the \fei\ lines at 630.15~nm,
630.25~nm, 1564.85~nm and 1565.29~nm. The main properties of the
transitions corresponding to these lines, such as the central wavelength
($\lambda_0$), the transition term, the effective Land\'e factor
($g_{\rm eff}$), the excitation potentials of the lower level
($\chi_{\rm e}$), and the weighted logarithmic oscillator strengths
($\log g^* f$) are given in Table~\ref{feiprops}. We have used a
  value of $\epsilon=7.43$ for the iron abundance
  \citep[e.g.,][]{shchukinabueno, Bellot-Rubio:Borrero:2002,
    Asplund:etal:2000b}. The $\log(g^* f)$ values for the infrared lines
  have been taken from \citet{Borrero:etal:2003}. For the 630.15~nm line
  we use the value given by \citet{Bard:etal:1991} while for 630.25~nm
  we use the same value as previous authors
  \citep[e.g.][]{Cabrera:etal:2005, Khomenko:etal:2005a,
    Khomenko:etal:2005b, Borrero:etal:2006}.

\begin{table}[hbt]
\begin{center}
\begin{tabular}{|c|c|c|c|c|} \hline
$\lambda_0$~[nm] & Transition & $g_{\mathrm{eff}}$ & $\chi_{\rm e}~\mathrm{[eV]}$ & $\log(g^* f)$ \\ \hline
630.15 & $ z^5P^0_2-e^5D_2 $ & 1.66 & 3.654 & -0.718 \\
630.25 & $ z^5P_1^0-e^5D_0 $ & 2.5  & 3.686 & -1.235 \\
1564.85 & $ e^7D_1-3d^64s5p^7D_1^0 $ & 3    & 5.43 & -0.675 \\
1565.29 & $ f^7D_5-(9/2)[7/2]^0_4 $ & 1.53 & 6.25 & -0.043 \\ \hline
\end{tabular}
\end{center}
\caption{Parameters of the spectral lines considered in this study.}
\label{feiprops}
\end{table}

The line profile calculations have been carried out by STOPRO routine
\citep{Solanki:1987, Frutiger:2000}.  It computes the full Stokes vector
by integrating the radiative transfer equations under the assumption of
local thermodynamic equilibrium (LTE), including Zeeman splitting and
magneto-optical effects. In order to obtain quantities required for line
profile calculations which are not delivered directly by the MURaM
result files, the MODCON routine was used. It determines the LTE
ionization equilibrium for a given chemical composition and derives
electron pressure, continuum optical depth, and continuum absorption
coefficient.

We have developed the program LINE, which calculates the Stokes
parameters for the 3D atmosphere simulated by the MURaM code. The
program uses STOPRO and MODCON as subroutines.  The program was
parallelized for computational efficiency of the line profile
calculations in large 3D computational domains.

\section{Results of Stokes diagnostics}
\subsection{General properties}

We have calculated Stokes profiles for both pairs of \fei\ lines, taking
vertical lines of sight through each of the $288\times288$ horizontal
resolution elements of the computational box.

In order to demonstrate the general consistency of synthetic line
profiles from simulations with observations, we show in
Figure~\ref{avcomp053000} a comparison of the spatially averaged
Stokes-$I$ profiles of the \fei\ line pairs in the infrared (upper
panel) and in the (lower panel) from the \bzavdec\ simulation (diamond
symbols) with the observed spectrum of the quiet Sun \citep[][black
curves]{delbouille}. Observed and simulated line profiles agree to
within a few percent. A more detailed comparison of mean profiles
would require considering a large number of snapshots from the
simulation run covering several periods of the 5-minute p-mode
oscillation, which is beyond the scope of this paper \citep[see,
e.g.,][]{Asplund:etal:2000a}.

In what follows, we shall concentrate on the results for one line in the
visible (630.25~nm, for simpler notation abbreviated to 630.2~nm in what
follows) and one line in the infrared (1564.85~nm, abbreviated to
1564.8~nm). We define a number of useful quantities to describe the
properties of the synthetic profiles, namely, the {\em line strength},
\begin{equation}
S=\frac{1}{\lambda_0} \int \frac{\left( I_c-I \right)}{I_c} {\mathrm d}
\lambda, \label{depreq}
\end{equation}
where $I$ is the (wavelength-dependent) intensity in the line, $I_c$ the
continuum intensity, and $\lambda_0$ the wavelength of the line center
in the rest frame.  It is clear from the definition that the line
strength is equal to the equivalent width normalized by the central
wavelength. Furthermore, we consider the {\em unsigned Stokes-$V$area,}
defined as
\begin{equation}
A_V=\int \frac{|V|}{I_c} {\mathrm d} \lambda, \label{stvareq}
\end{equation}
where $V$ is the Stokes $V$ parameter, and the {\em Stokes-$V$area
asymmetry}\footnote{The sign of the net magnetic flux through the
simulation box has been chosen such that, for a normal two-lobed
$V$-profile, $\delta A_V$ is equal to the result of the usual definition
$(|A_b|-|A_r|)/(|A_b|+|A_r|)$, where $A_b$ and $A_r$ are the areas of
the `blue' and `red' lobes, respectively
\citep[e.g.,][]{Solanki:Stenflo:1984}.}
\begin{equation}
\delta A_V=\frac{\int V{\mathrm d} \lambda}{\int |V| {\mathrm d} \lambda}, 
\label{stvaseq}
\end{equation}
as well as the {\em Stokes-$V$amplitude asymmetry,} defined as
\begin{equation}
\label{stvaseqa}
\delta a_V=(|a_b|-|a_r|)/(|a_b|+|a_r|),
\end{equation}
where $a_b$ is the amplitude of the `blue' lobe of the Stokes-$V$ profile
and $a_r$ is the corresponding quantity of the `red' lobe. In all cases
above, the integrals are taken over the full width of the line.

\begin{figure*}
\centering
\includegraphics[width=\hsize]{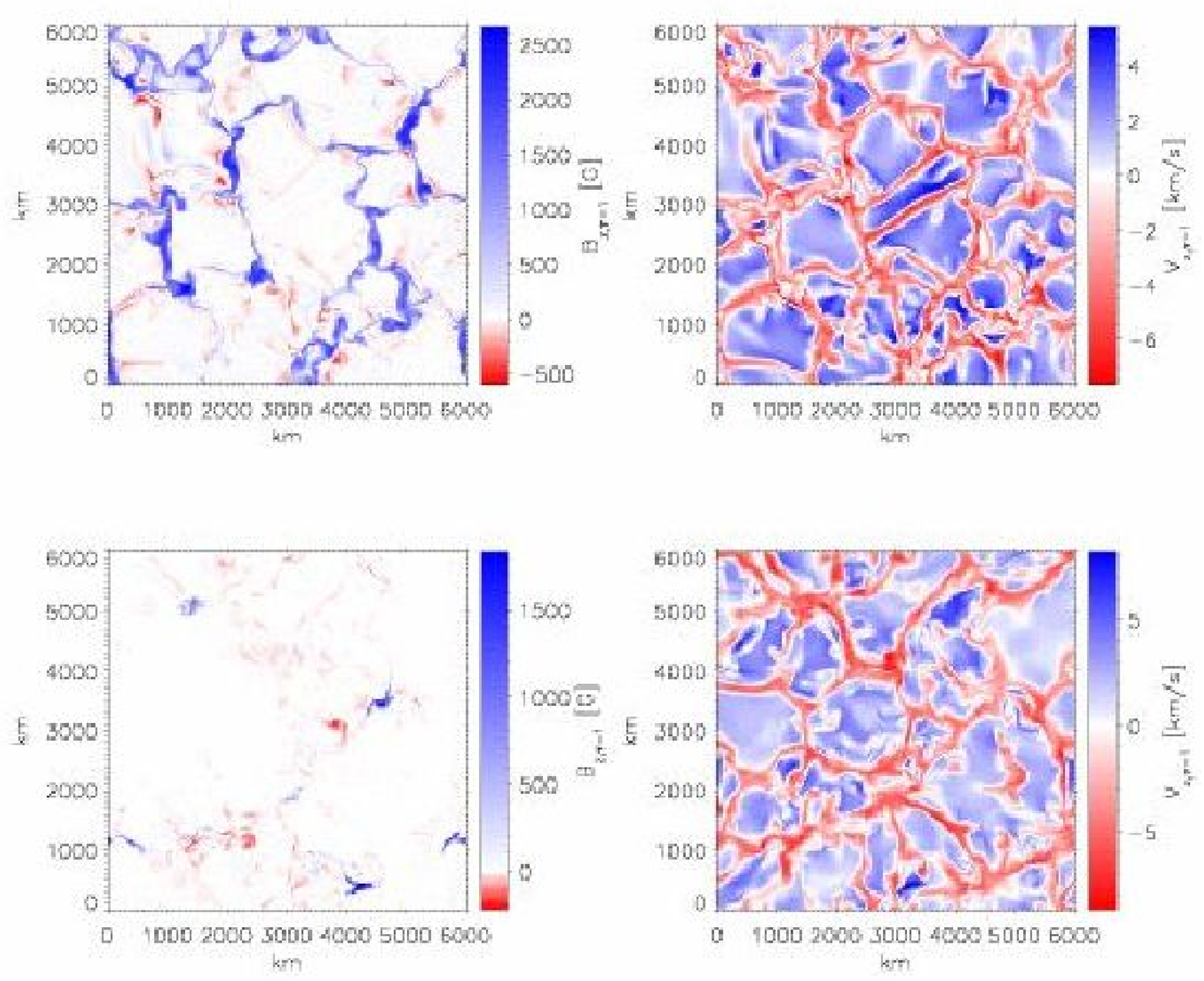}
\caption{Vertical magnetic field component (left panels) and vertical
velocity (right panels) on the surface $\tau_{500}=1$ (continuum optical
depth unity at 500 nm wavelength) for snapshots from simulations with
\bzavpl\ (plage, upper row) and \bzavdec\ (decay phase, bottom row).
Positive values of the velocity (shown in blue) correspond to upflows;
downflows are are indicated in red. Note the different scales for
positive and negative $B_z$ in the left panel (all negative-polarity
field is rather weak).}
\label{simpic}
\end{figure*}

\begin{figure}
\includegraphics[width=\hsize]{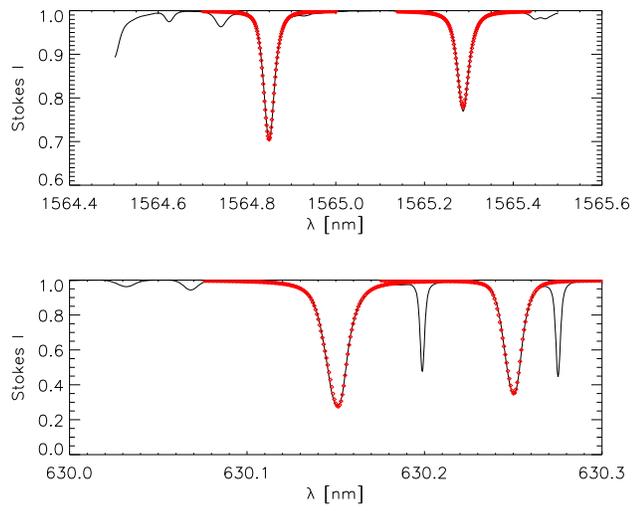}
\caption{Comparison between the spatially averaged synthetic \fei\ line
profiles for the simulation with \bzavdec\ (diamond symbols) and
observed quiet-Sun profiles (solid curves) from the spectral atlas of
\citet{delbouille}.}
\label{avcomp053000}
\end{figure}

\begin{figure*}
\centering
\includegraphics[width=\hsize]{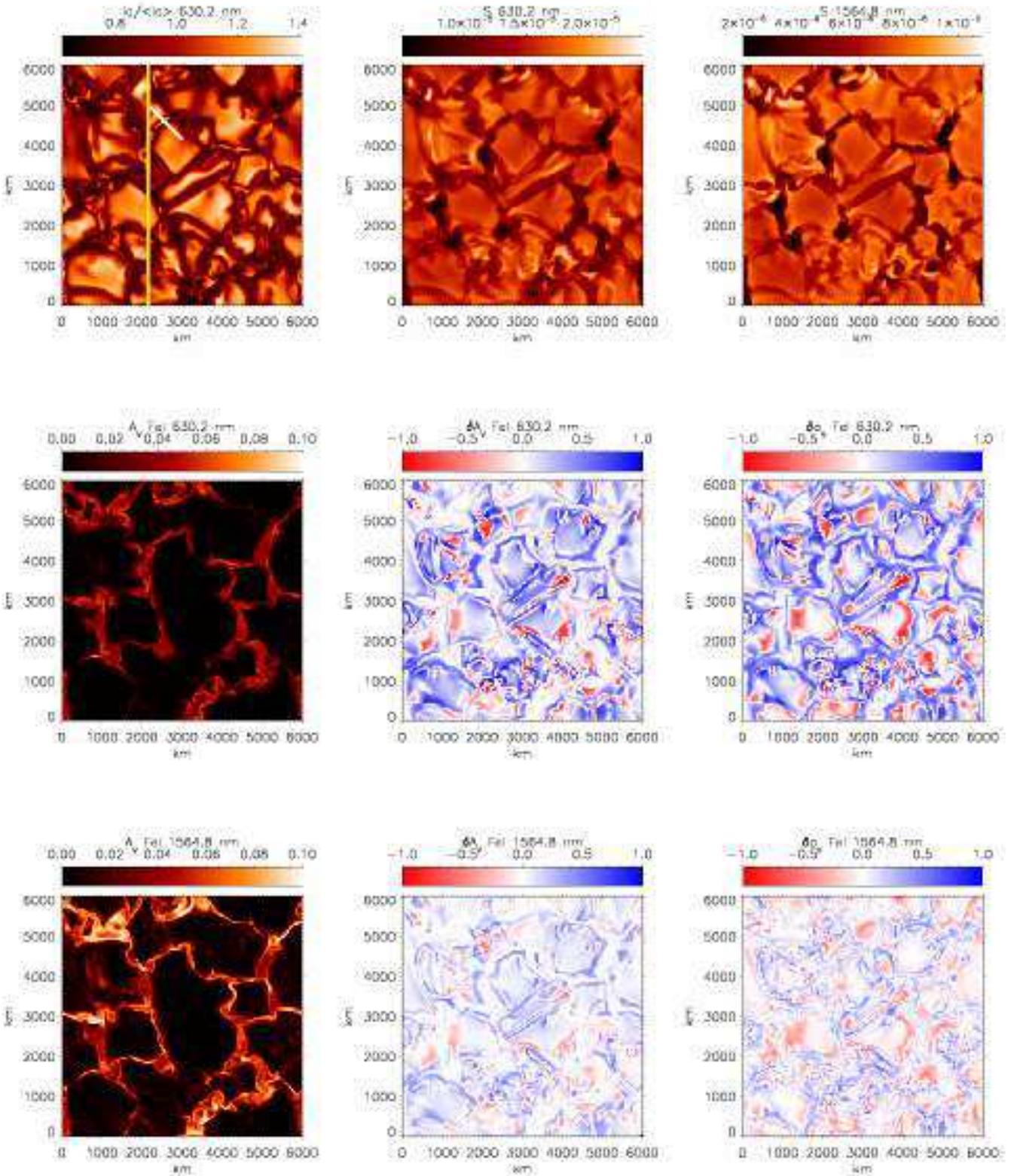}
\caption{Properties of synthetic Stokes-$I$ and Stokes-$V$profiles for
the snapshot from the \bzavpl\ simulation shown in
Fig.~\ref{simpic}. Given are maps of the (normalized) continuum
intensity at 630.2~nm (top left panel), line strength, $S$ (top middle
panel for \fei\ 630.2~nm and top right panel for \fei\ 1564.8~nm), unsigned
Stokes-$V$ area, $A_V$ (middle left panel for 630.2~nm and bottom left
panel for 1564.8~nm), Stokes-$V$ area asymmetry (central panel for
630.2~nm and bottom middle panel for 1564.8~nm), and Stokes-$V$ amplitude
asymmetry (middle right panel for 630.2~nm and bottom right panel for
1564.8~nm). In the continuum intensity image (top left panel), the slit
position for Figs.~\ref{slit6302} and \ref{slit15648} is indicated by
the vertical line and the cut through the magnetic flux sheet shown in
Fig.~\ref{flt2dcut} is indicated by the inclined line in the upper
left quadrant.}
\label{images088200}
\end{figure*}

Figure~\ref{images088200} shows some results obtained for the snapshot
from the \bzavpl\ (plage) run. The normalized continuum image for
$\lambda=630.2\,$nm shows localized brightness enhancements
corresponding to the concentrations of strong vertical magnetic
field. The latter are clearly detectable in the maps of the unsigned
Stokes-$V$ area (middle left panel for the 630.2~nm line, bottom left
panel for 1564.8~nm) as can be seen by comparing with the magnetic field
map in the top left panel of Fig.~\ref{simpic}. While the unsigned
Stokes-$V$ area in both lines follows the total field strength rather
well, the larger Zeeman sensitivity of the 1564.8~nm infrared line is
clearly reflected in the maps, more than compensating for its smaller
line strength. The larger $A_V$ value of this line may partly also
result from its lower formation height and from the fact that strong
flux concentrations show a larger height gradient of the field strength
(compare the magnetic field maps in Fig.~\ref{simpic} with those in
Figs.~\ref{moments088200} and \ref{moments053000}).  The corresponding
maps of the line strength (top middle and top right panels,
respectively) reveal significant line weakening in the regions of strong
magnetic field.  The maps of the Stokes-$V$ area asymmetry (central
panel and bottom middle panel, respectively) and of the amplitude
asymmetry (middle right and bottom right panel, respectively) show
strong asymmetries in the periphery of magnetic flux concentrations,
while their central parts exhibit almost no asymmetry.  The
predominantly positive values indicate that in most cases the `blue'
lobe the of Stokes-$V$ profile has larger area and amplitude than the
`red' lobe.  Lines of sight through the peripheral parts of a flux
concentration that expands with height cross its boundary and, in the
presence of strong external flows, thus sample significant magnetic
field and velocity gradients. As a result, the Stokes-$V$ profiles
become asymmetric \citep{illingetal} with a stronger blue lobe in the
case of dominating external downflows
\citep{Grossmann-Doerth:etal:1988}. The asymmetry arises from the fact
that the wavelength shifts due to the Doppler effect and the Zeeman
splitting have the same sign for one lobe, but opposite sign for the
other lobe. The effect is strongest when both shifts are of similar
magnitude; in addition, the area asymmetry depends on line saturation
\citep{Grossmann-Doerth:etal:1989,Solanki:1989}. This explains why the
asymmetries are considerably larger for the visible line, which is
stronger (more saturated) and has a smaller Zeeman splitting, better
matching the Doppler shift. On the other hand, the spatial distributions
of $\delta A_V$ and $\delta a_V$ from both lines are very similar.

\begin{figure*}
\centering
\includegraphics[width=\hsize]{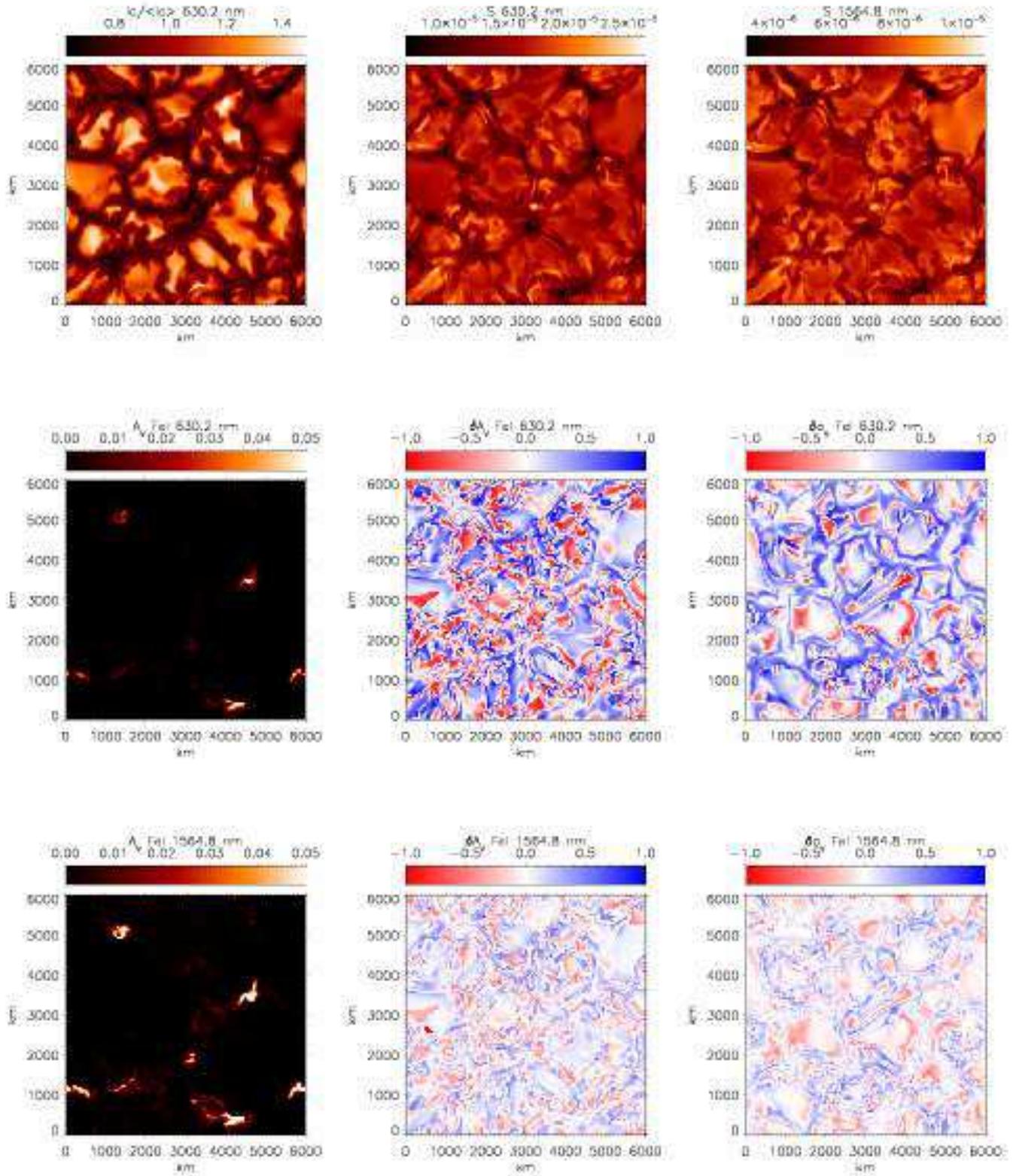}
\caption{Same as Fig.~\ref{images088200} for the snapshot from the
  simulation with \bzavdec . The scales are the same as in the previous
  figure, except for the $A_V$ images.}
\label{images053000}
\end{figure*}

Figure~\ref{images053000} shows the corresponding maps for the \bzavdec\
run, which could represent an almost decayed unipolar part of an active
region or the field underlying a coronal hole.  The maps of unsigned
Stokes-$V$ area (middle left and bottom left panels) show a significant
signal only in a few isolated patches of concentrated magnetic flux.
The infrared line displays an enhanced $A_V$ not only in the weak-field
regions, but also in the stromg flux concentrations. This does not imply
that the Stokes-$V$ amplitude is also larger for this line, since the
$V$ lobes can become very broad owing to the vertical field gradient
\citep{Zayer:etal:1989}. The maps of the Stokes-$V$ area and amplitude
asymmetries are very complex and show almost no correspondence to the
structures in the continuum image. In this case, the strongest
asymmetries arise from very weak magnetic fields in granules, with some
preference for the granule edges. No sign of the asymmetry parameters
is preferred, with the 630.2~nm showing much larger values.

\begin{figure*}
\centering
\includegraphics[width=\hsize]{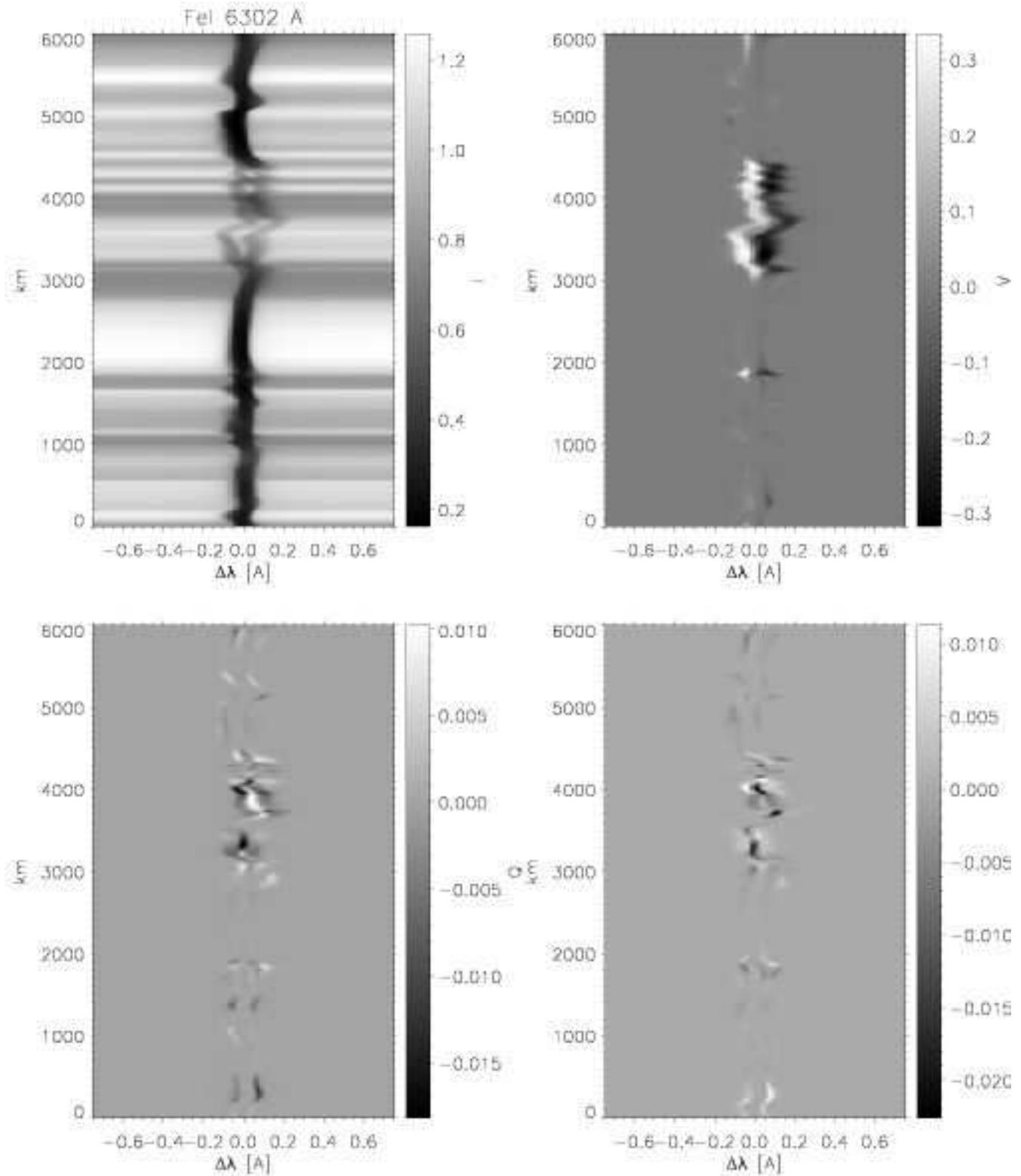}
\caption{Synthetic slit spectra, calculated for the 630.2~nm line. Top
left: Stokes-$I$, top right: Stokes-$V/I$, bottom left: Stokes-$Q/I$,
bottom right: Stokes-$U/I$. The position of the slit is indicated by the
vertical line in the top left panel of Fig.~\ref{images088200}.}
\label{slit15648}
\end{figure*}

\begin{figure*}
\includegraphics[width=\hsize]{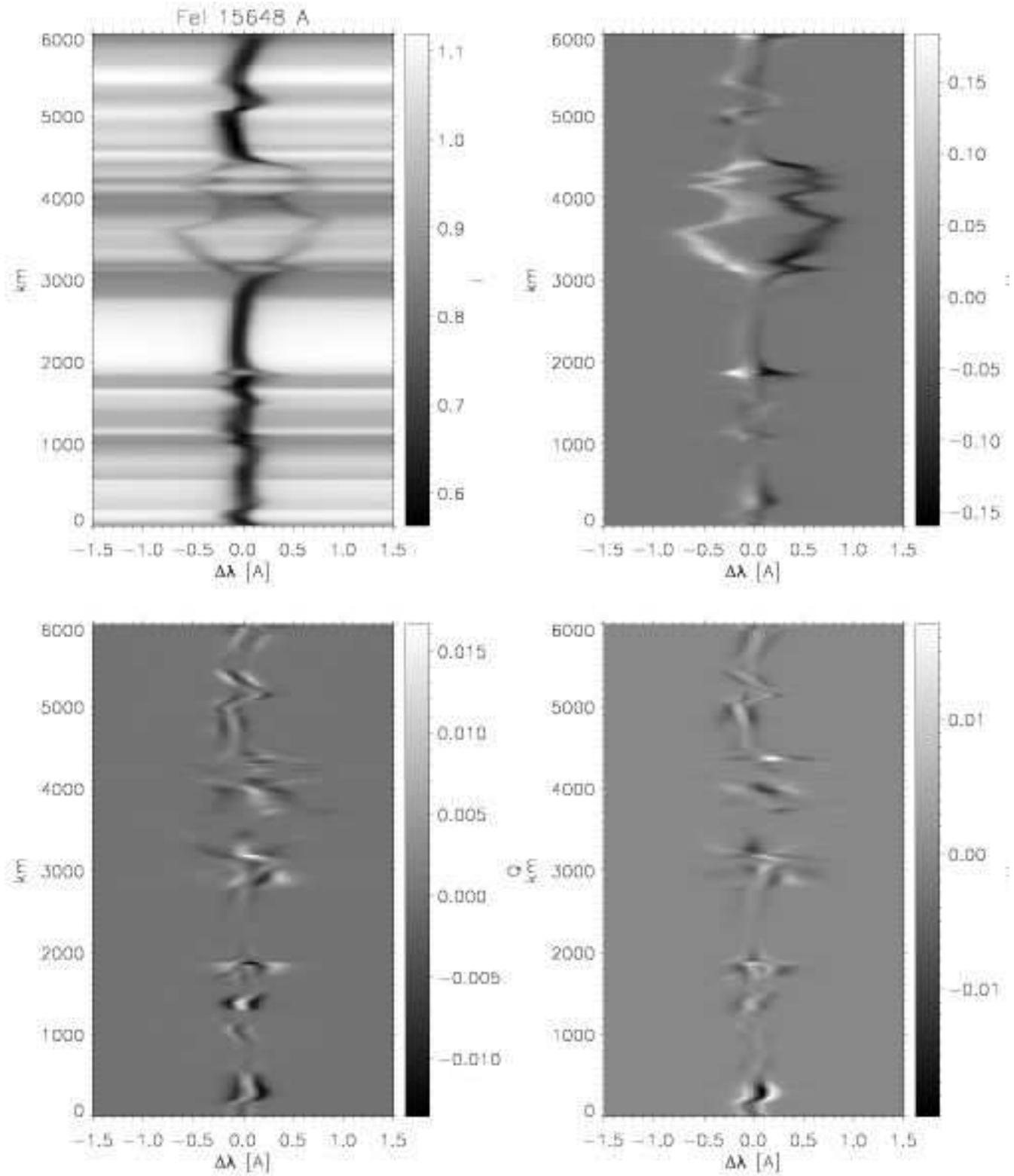}
\caption{Same as Fig.~\ref{slit15648} for the 1564.8~nm line.} 
\label{slit6302}
\end{figure*}

\begin{figure*}
\centering
\includegraphics[width=\hsize]{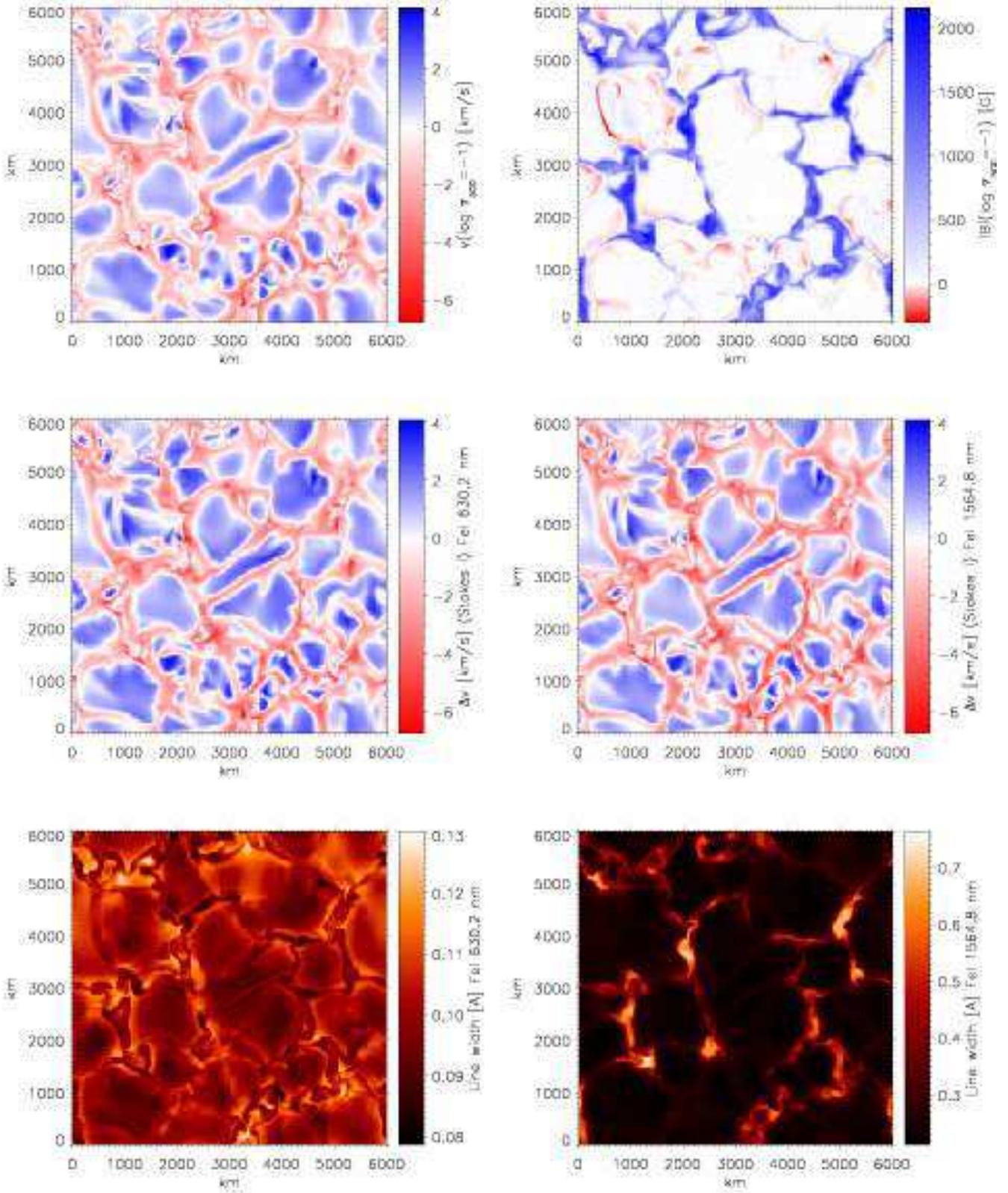}
\caption{Vertical velocity (top left panel) and field strength (top
right panel) at the surface $\tau_{500}=0.1$ for the \bzavpl\
simulation. The middle row shows maps of the vertical velocity as
determined from the Doppler shift of the 630.2~nm line (middle left
panel) and the 1564.8~nm line (middle right panel). The bottom row gives
maps of the line width of the 630.2~nm line (bottom left panel) and the
1564.8~nm line (bottom right panel).}
\label{moments088200}
\end{figure*}

\begin{figure*}
\centering
\includegraphics[width=\hsize]{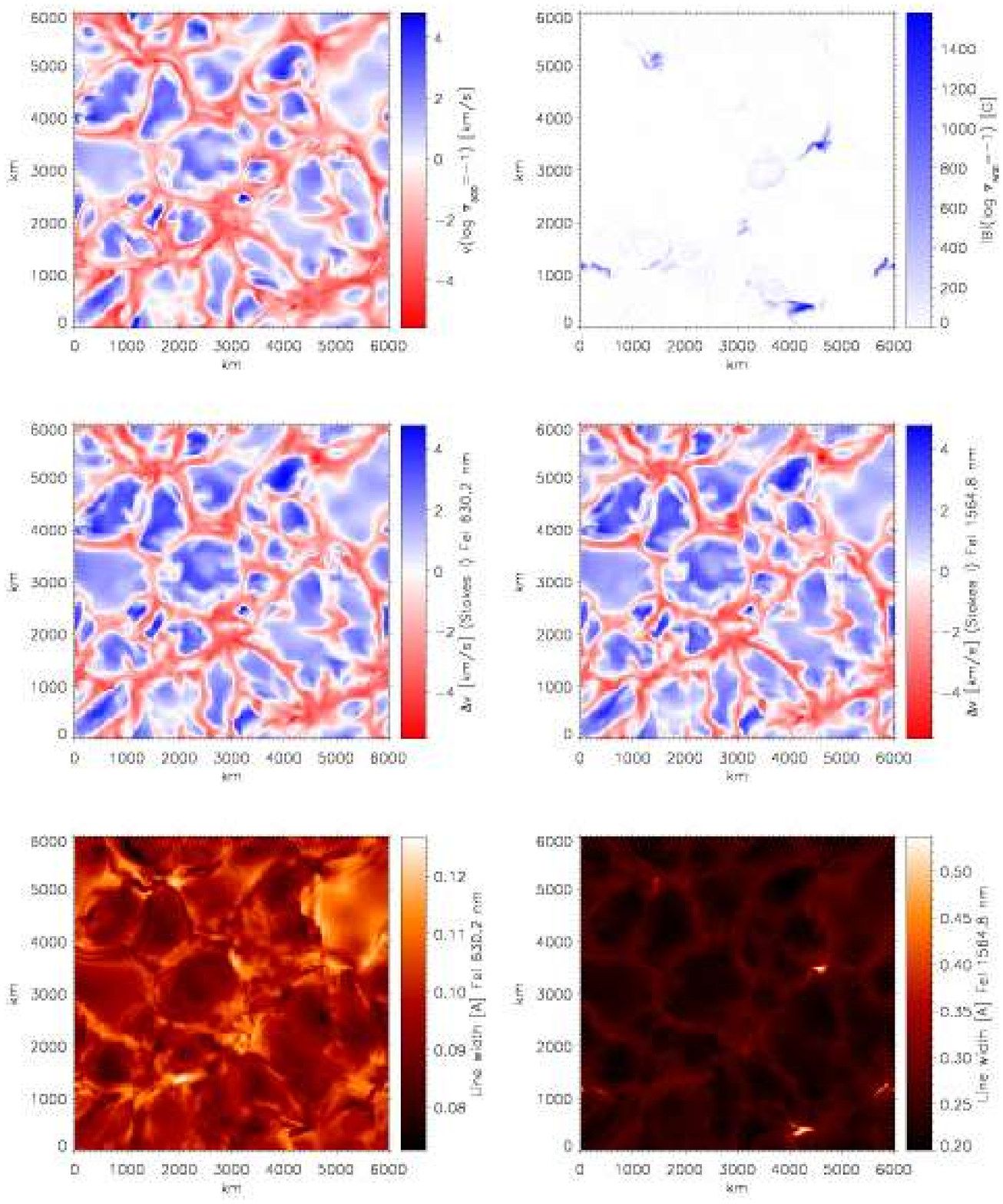}
\caption{Same as Fig.~\ref{moments088200} for the \bzavdec\ simulation.}
\label{moments053000}
\end{figure*}

Synthetic `slit spectra' of all four Stokes parameters along the
vertical yellow line on the continuum image for the 200~G snapshot
(upper left panel of Fig.~\ref{images088200}), are shown in
Figs.~\ref{slit15648} and \ref{slit6302}. The width of the
artificial slit is equal to the horizontal cell size of the simulation
(about 20 km on the Sun). The spectra represent `ideal' cases; for a
direct comparison with actual observations, one would have to take into
account the instrumental profile and finite spatial resolution.
The synthetic spectral lines show the characteristic `line wiggles' from
the Doppler shifts due to the granular up- and downflows.  The
Stokes-$V$ spectra (upper right panel) show strong signals where the
slit cuts through an extended flux concentration (in the upper half) and
through a narrow flux sheet (at $\sim 1800~\mathrm{km}$).  The
$I$-profiles of the lines are weakened and split (more strongly so in
the case of the infrared line) at these locations, while strong line
wiggles in both Stokes-$I$ and Stokes-$V$ indicate the presence of
significant internal structure of the large magnetic flux
concentration. Note that some of the variation may be due to the fact
that the large, elongated flux sheet is bent, so that central and
peripheral regions are sampled at different locations along the
slit. The spectra of Stokes-$Q$ and $U$ show rather weak signals,
consistent with the lack of significant horizontal magnetic field
components in the strong flux concentrations. As expected from its
larger Zeeman sensitivity, the infrared line displays larger values of
$Q$ and $U$. .

In order to compare spectroscopically determined velocities with their
actual values in the simulation, we define the line-of-sight velocity,
$\Delta v$, according to the Doppler shift of the Stokes-$I$ profile,
viz.
\begin{equation}
\Delta v=\frac{c(\lambda_0-\lambda_{\rm cg})}{\lambda_0},
\label{deltaveq}
\end{equation}
where $c$ is the velocity of light, $\lambda_0$ is the wavelength of the
line center in the rest frame, and $\lambda_{\rm cg}$ is the wavelength
corresponding to the center of gravity of the line. The latter is
defined as the first moment of the profile,
\begin{equation}
\lambda_{\rm cg}=\frac{\int (I_c-I) \lambda d\lambda}{\int (I_c-I)
d\lambda}, \label{lceq}
\end{equation}
where $I_c$ is the continuum intensity and $I$ is the line intensity. 

\begin{figure*}
\centering
\includegraphics[width=\hsize]{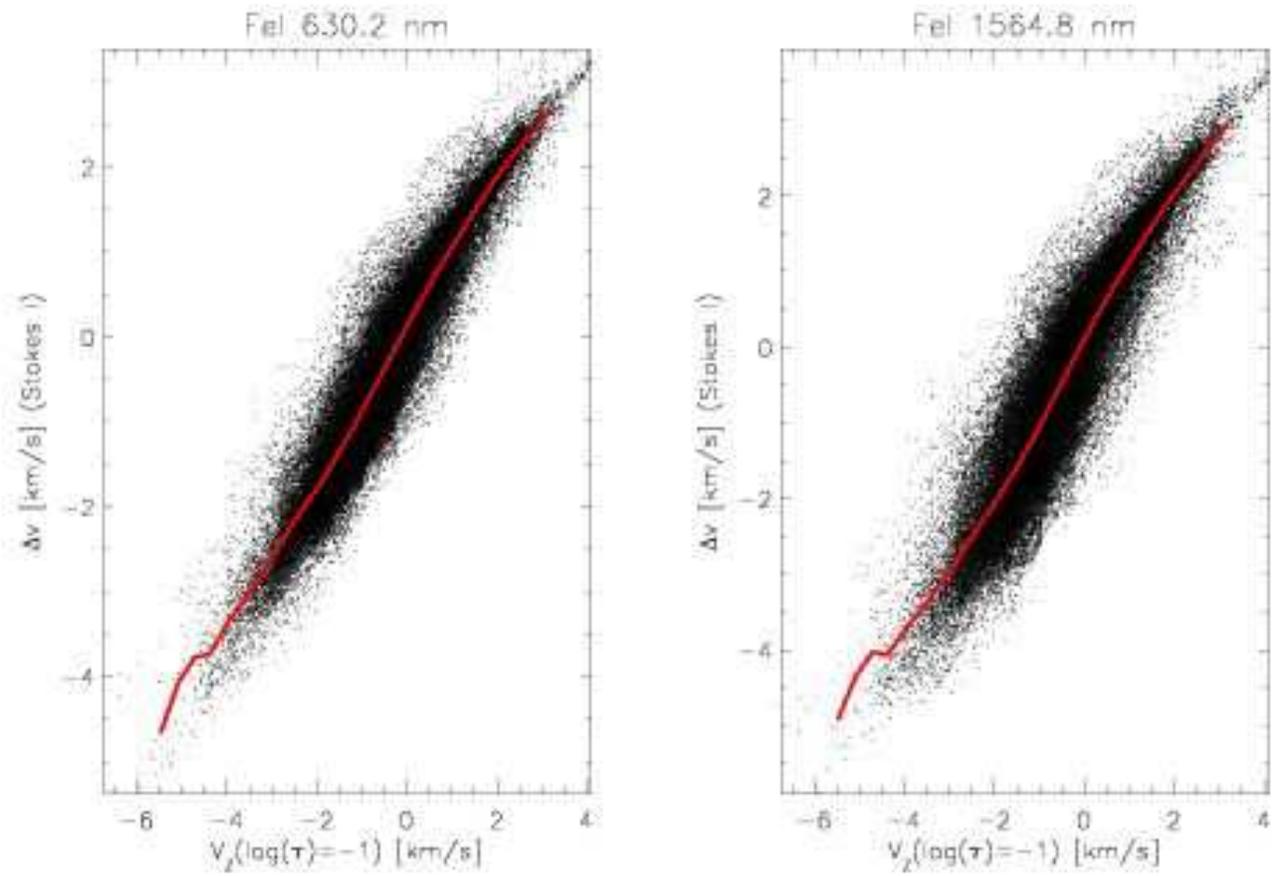}
\caption{Scatter plots of velocity determined from the Doppler shift of
  the center of gravity of the spectral line versus the vertical
  velocity at $\tau_{500}=0.1$ from the simulation (200 G). The line
  shows the binned average.}
\label{velcompar}
\end{figure*}

Fig.~\ref{moments088200} (for the \bzavpl\ simulation) and
Fig.~\ref{moments053000} (for \bzavdec) give a comparison between the
velocity $\Delta v$ determined from the Doppler shift of the 630.2~nm
(middle left panel) and 1564.8~nm lines (middle right panel),
respectively, with the actual flow pattern of the simulation at the
surface $\tau_{500}=0.1$ (top left panel). This level can be seen as a
very rough representation of the optical depth range contributing to the
spectral line shift. The maps show that the actual velocities at that
height and the spectroscopically determined velocities qualitatively
agree. The somewhat larger flow velocities returned by the infrared line
are consistent with its lower height of formation since the flow speeds
tend to increase with depth. However, the corresponding scatter plots
given in Fig.~\ref{velcompar} show that the scatter of the individual
points is considerable. This is not surprising in view of the extended
height range of line formation (or better, the finite width of the
velocity response function in combination with significant vertical
velocity gradients). In fact, it has been shown that a much better
representation of the height to which the center-of-gravity shift of
Stokes-$I$ corresponds can be determined via response
functions \citep[][Sect. 10.4.3]{Beckers:Milkey:1975, Landi:Landi:1977,
Grossmann-Doerth:etal:1988c, DelToroIniesta:etal:1994,
Sanchez:etal:1996, Cabrera:etal:2005, DelToro:2003}. Nevertheless, it is
interesting to see that, for a first guess, the level
$\tau_{500}=0.1$ is not a completely unreasonable choice.

The bottom panels of Figs.~\ref{moments088200} and \ref{moments053000}
show maps of the line width, $\Delta\lambda$, which is defined here as
the standard deviation of the profile (square root of the second
moment):
\begin{equation}
\Delta\lambda=\sqrt{\frac{\int(I_c-I)(\lambda-\lambda_c)^2
d\lambda}{\int(I_c-I)d\lambda}}. \label{lweq}
\end{equation}
The maps of $\Delta\lambda$ (bottom left panel for 630.2~nm,
bottom right panel for 1564.8~nm) reflect the different magnetic
sensitivity of the visible and infrared \fei\ lines. In the case of
630.2~nm, the line width is actually decreased in the bulk of the
magnetic flux concentrations, showing that its Zeeman sensitivity is
insufficient for a significant broadening. On the other hand, there is
a clear increase of $\Delta\lambda$ in the periphery of the flux
concentrations, where the field is weaker but strong downflows
prevail. In connection with the fact that the line width is found to
be also enhanced in non-magnetic intergranular downflow lanes (see
Fig.~\ref{moments053000}), this indicates that the strong velocity
gradients in downflows are responsible for the broadening of the
visible line \citep{nesisgran,solankigran}.  In the case of the more
Zeeman-sensitive 1564.8~nm infrared line, the maps show a clear
correspondence between the magnetic field strength and the line
width. Doppler broadening is relevant for this line as well, since
some enhancement of the line width can be seen in all downflow lanes,
even in the absence of a strong field.

\subsection{Cut through a magnetic feature}

\begin{figure*}
\centering
\includegraphics[width=\hsize]{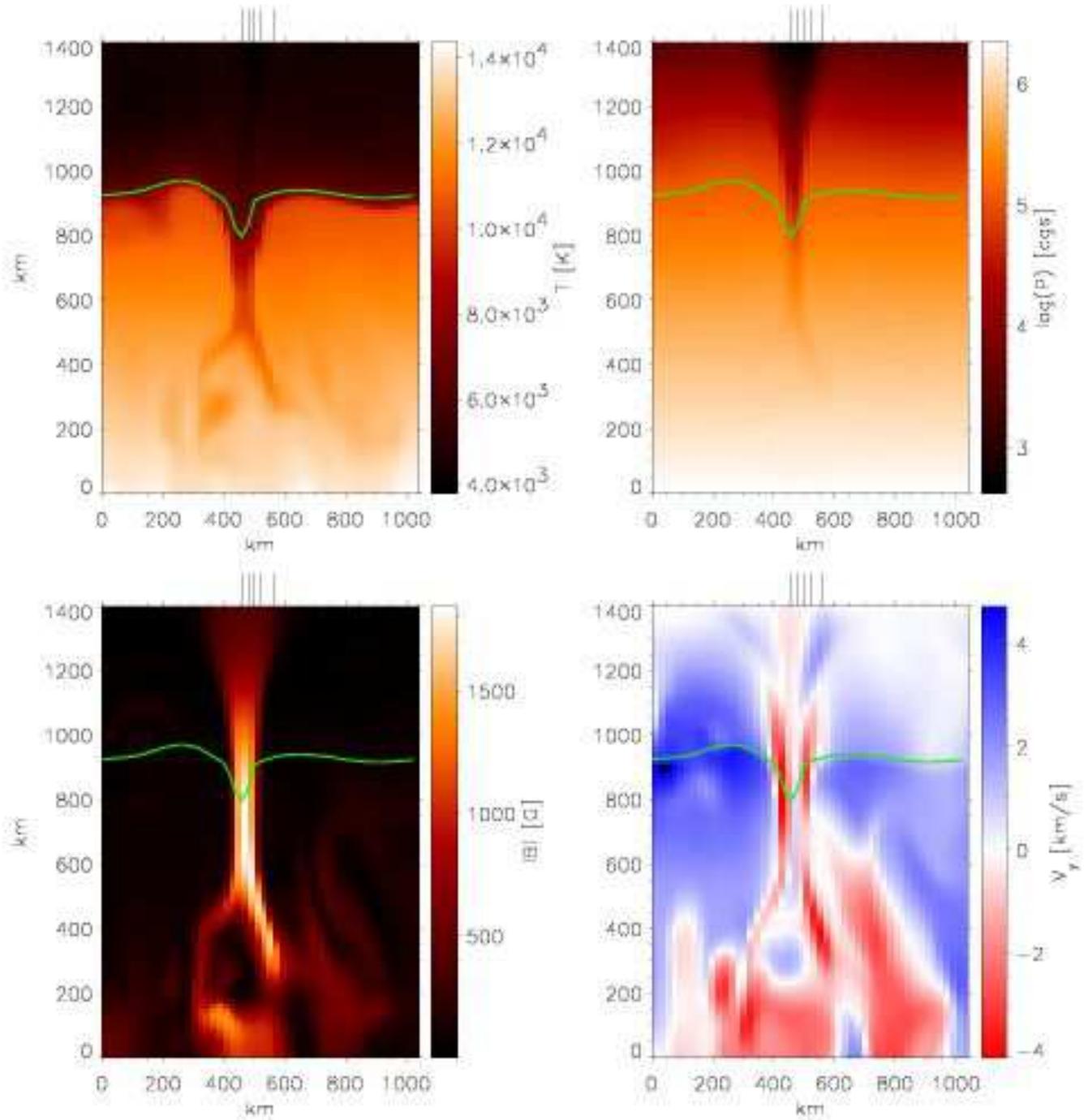}
\caption{Temperature (upper left panel), gas pressure (upper right
panel), magnetic field strength (lower left panel), and vertical
velocity (lower right panel) for the vertical cut through the simulation
box marked by the white line on the upper left panel of
Fig.~\ref{images088200}. The green curve indicates the level
$\tau_{500}=1$. Negative values of the velocity (shown in red)
correspond to downflows. The lines of sight corresponding to the Stokes
profiles and gradients shown in Fig.~\ref{088200prof} are indicated at
the top of each panel.}
\label{flt2dcut}
\end{figure*}

A two-dimensional vertical cut through a sheet-like magnetic feature
along the inclined white line in the upper left panel of
Fig.~\ref{images088200} is shown in Fig.~\ref{flt2dcut}. The four panels
display the distributions of temperature (top left), gas pressure (top
right), magnetic field strength (bottom left), and vertical velocity
(bottom right). The green line on the images indicates the
$\tau_{500}=1$ level. The analyzed flux sheet is rather shallow, the
flux concentration extending only to a depth of about 500~km below the
average level of $\tau_{500}=1.$ With respect to its surroundings at the
same geometrical depth, the gas in the interior of the magnetic flux
sheet is cooler and has a deficit in gas pressure, which is almost
completely compensated by the magnetic pressure, in accordance with the
approximation of thin flux tubes \citep{Shelyag:2004}. There are strong
downflows with velocities up to 3~km$\,$s$^{-1}$ at the periphery of the
magnetic flux sheet, while the flow speeds in its interior are smaller,
exhibiting an upflow of up to 600~m$\,$s$^{-1}$ near $\tau_{500}=1$.
Note that the velocity plot in Fig.~\ref{flt2dcut} shows only the
vertical component: since the flow is largely subsonic, a divergence of
the vertical mass flux at its stagnation points is balanced by a
convergence of the horizontal mass flux, and vice versa.
The height level of $\tau_{500}=1$ is depressed by about 150~km in the
region of strong magnetic field. This is caused by the lower gas
temperature and density in the flux sheet, an effect analogous to the
Wilson depression in sunspots.

\begin{figure*}
\centering
\includegraphics[width=\hsize]{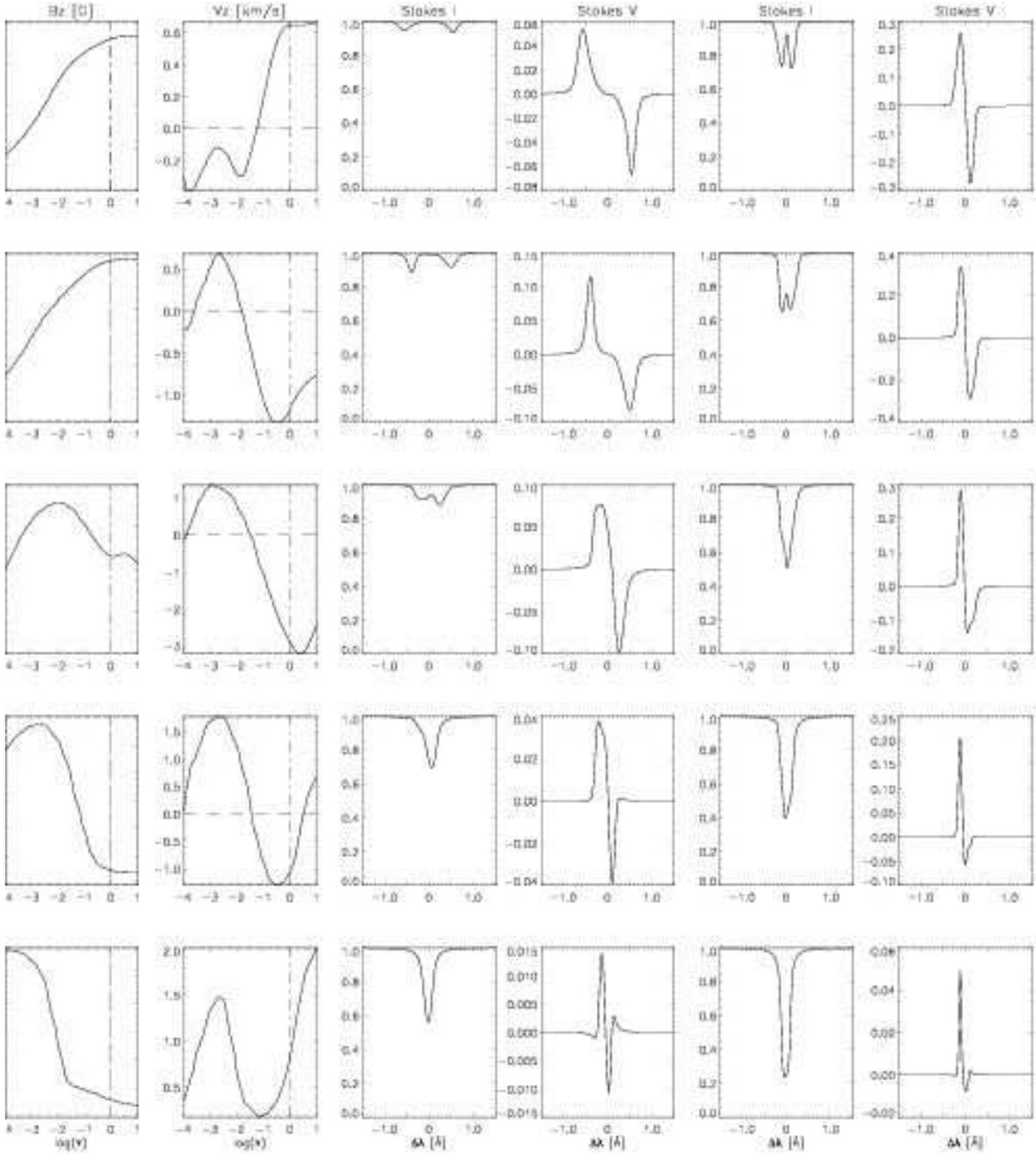}
\caption{Stokes-$I$ and Stokes-$V$ profiles for the 1564.8~nm line (third
  and fourth column) and for the  630.2~nm line (fifth and sixth
  column). The corresponding magnetic field and velocity profiles as a
  function of $\log\tau_{500}$ are given in the first two columns. The level
  $\tau_{500}=1$ is marked by the dash-dotted line; zero vertical
  velocity is indicated by the dashed line. Each row corresponds to one
  of the vertical lines of sight indicated in Fig.~\ref{flt2dcut},
  ranging from the central part of the flux concentration (top row) to
  its outer periphery (bottom row). }
\label{088200prof}
\end{figure*}

Fig.~\ref{088200prof} shows the profiles of magnetic field and velocity
along several vertical lines of sight (indicated by the short vertical
lines at the top of the panels in Fig.~\ref{flt2dcut}) together with the
emergent Stokes-$I$ and Stokes-$V$ profiles at these locations. Each
line of sight corresponds to one row of panels in Fig.~\ref{088200prof},
which show (from left to right) the line-of-sight components of the
magnetic field and the velocity field as a function of $\tau_{500}$, and
the Stokes-$I$ and $V$ profiles of the 1564.8~nm and 630.2~nm lines.
The first row in Fig.~\ref{088200prof} represents the center of the flux
sheet, showing a slow upflow and large field strength in the height
range relevant for line formation, so that the Stokes profiles are split
and nearly symmetric. The field strength drops with height in accordance
with the horizontal pressure balance.  Moving towards the periphery of
the flux concentration, the lines of sight cut through the magnetopause
of the expanding magnetic structure into the surrounding downflow
(cf. the last three rows of Fig.~\ref{088200prof}, which show an
increase of the field strength with height in the lower photosphere). As
a result, the Zeeman splitting is smaller and the $V$-profiles,
particularly for the visible line, become strongly asymmetric.

\citet{illingetal} have shown that asymmetric Stokes-$V$ profiles arise
in the presence of magnetic field and velocity gradients along the line
of sight, the sign of the asymmetry being determined by the sign of the
product of both gradients.  On this basis, a mechanism leading to
strongly asymmetric (and unshifted) Stokes-$V$ profiles in the periphery
of a static magnetic flux concentration has been suggested by
\citet{grosschsol} and \citet{Solanki:1989}. These authors assumed an
idealized two-component atmosphere model: the upper part of the
atmosphere has a magnetic field and is static while the lower part is
non-magnetic and exhibits a systematic vertical flow. The Stokes-$V$
profile formed along a vertical line of sight that cuts through both
components then becomes asymmetric with an unshifted zero-crossing
wavelength.  \citet{solankipahlke} showed that the blue wing of the
Stokes-$V$ profile is stronger than the red wing (positive area
asymmetry) if the inequality
\begin{equation}
\frac{{\mathrm d}|B_\mathrm{los}|}{{\mathrm d}\tau} 
\cdot\frac{{\mathrm d}v_\mathrm{los}}{{\mathrm d} \tau}>0
\label{signaseq}
\end{equation}
holds, where $B_\mathrm{los}$ and $v_\mathrm{los}$ are the line-of-sight
components of the magnetic field and the flow velocity,
respectively. Note that positive values of the velocity correspond to
upflows.  The asymmetry is positive in regions where the velocity and
the magnetic field gradients have the same sign, such as in the
periphery of the flux sheet shown in Figs.~\ref{flt2dcut} and
\ref{088200prof}. The area asymmetries of the $V$-profiles corresponding
to the five lines of sight considered in Fig.~\ref{088200prof} are in
agreement with this rule (compare Table~\ref{asytab6302} with the first
two columns of Fig.~\ref{088200prof}). Note that the Stokes-$V$
profiles are formed higher in the atmosphere near the periphery of the
flux tube, since the they obtain a larger contribution from layers with a 
stronger field.
The fact that the 630.2~nm line generally
shows larger $V$-profile asymmetry than 1564.8~nm line is due to the
combination of stronger saturation and weaker splitting of the visible
line \citep{Grossmann-Doerth:etal:1989,Solanki:1989}.

\begin{table}[hbt]
\begin{center}
\begin{tabular}{|c|r|r|r|r|} \hline
 & \multicolumn{2}{|c|}{630.2~nm} & \multicolumn{2}{c|}{1564.8~nm} \\ \hline
l.o.s. \# & $\delta a_V$ & $\delta A_V$ & $\delta a$ & $\delta A$ \\ \hline
1 & $-$0.02  &  0.03&    $-$0.11 &    0.01 \\
2 &    0.06  &  $-$0.07  &  0.16 &   $-$0.03  \\
3 &  0.31 &     0.05 &   $-$0.12 &  0.04  \\
4 &  0.51  &   0.43  &  $-$0.02  &  0.33  \\
5 &    0.72  &  0.59  &     0.13 &  0.27   \\ \hline
\end{tabular}
\end{center}
\caption{Stokes-$V$ amplitude asymmetry ($\delta a_V$) and area
asymmetry ($\delta A_V$) for the 630.2~nm and 1564.8~nm lines along 5
lines of sight (l.o.s.) in Figs.~\ref{flt2dcut} and
~\ref{088200prof}. The l.o.s. \# 1 is near the center of the flux sheet
while l.o.s. \# 5 is in its far periphery.}
\label{asytab6302}
\end{table}

\subsection{Statistical properties}

We now consider the complete set of Stokes profiles calculated for
vertical lines of sight through each of the $288^2$ horizontal
resolution elements of $20.8\times20.8\,$km$^2$ size each, which are
provided by a simulation snapshot.  We study the dependence of their
properties on the magnetic field and flow structure using the snapshot
from the \bzavpl\ simulation.

\begin{figure*}
\centering
\includegraphics[width=\hsize]{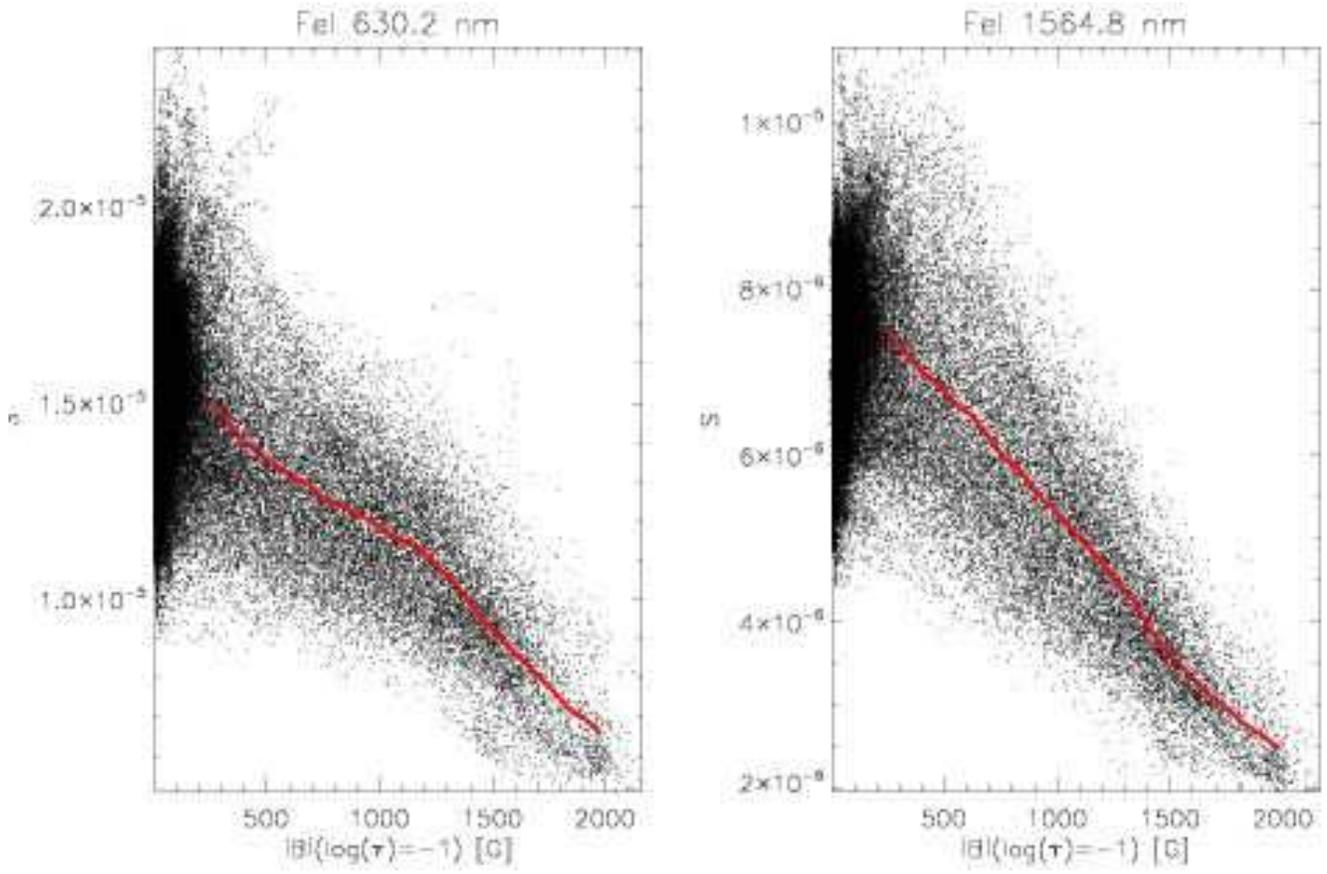}
\caption{Scatter plots of line strength versus magnetic field strength
at $\tau_{500}=0.1$ for the snapshot from the \bzavpl\ simulation. {\em
a:} \fei\ 630.2~nm, {\em b:} \fei\ 1564.8~nm.  The lines show the
binned averages. Note the different vertical scales on the two panels.}
\label{btaustokesi}
\end{figure*}

\begin{figure*}
\centering
\includegraphics[width=\hsize]{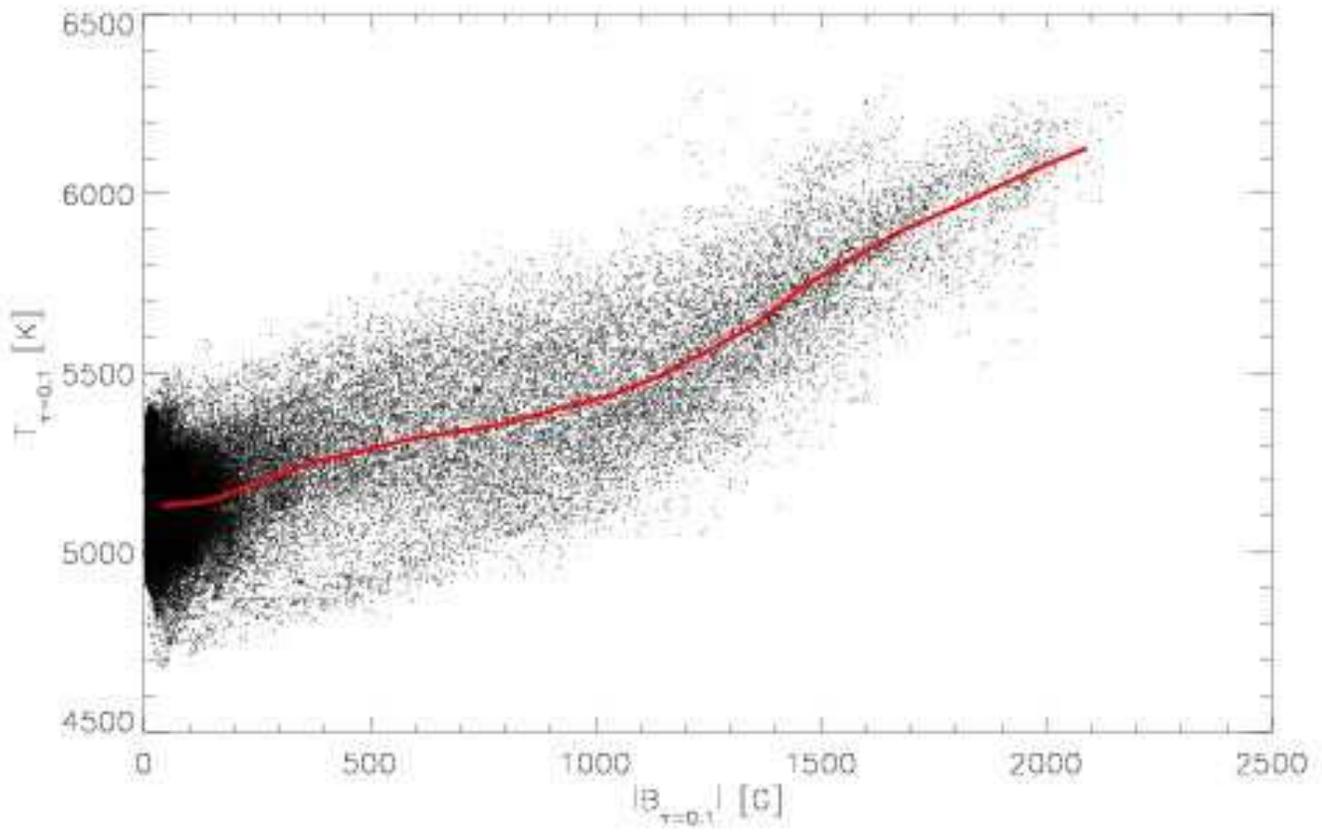}
\caption{Scatter plots of temperature versus magnetic field strength
(both at $\tau_{500}=0.1$).  The line shows the binned average.}
\label{temp}
\end{figure*}

Scatter plots of the line strength $S$ versus magnetic field strength at
$\tau_{500}=0.1$ for the 630.2~nm line (Fig.~\ref{btaustokesi}a) and for
the 1564.8~nm line (Fig.~\ref{btaustokesi}b) show that both lines are
progressively weakened with increasing magnetic field strength. The
binned averages given by the line indicate an almost linear
dependence, in the case of the 630.2~nm line with some increase of the
slope for large field strength.  The scatter plot of temperature versus
field strength (both at $\tau_{500}=0.1$) shown in Fig.~\ref{temp}
indicates that the line weakening is largely due to the larger
temperature (at equal optical depth) in the magnetic flux concentrations
\citep[cf.][]{Chapman:Sheeley:1968}, resulting from the lateral
radiative heating and reduced density of these structures. However, the
visible line, which should be more temperature sensitive owing to its
lower value of $\chi_{\mathrm e}$, shows a somewhat lower relative
weakening than the infrared line. This indicates that other effects,
such as Zeeman desaturation as well as vertical velocity and temperature
gradients, also affect the line strength in a significant way. A
more detailed consideration on the basis of response functions
\citep[e.g.,][Sect. 10.4.2]{DelToro:2003} is required to shed more light
on this result, but such an analysis is beyond the scope of this paper.

\begin{figure*}
\centering
\includegraphics[width=\hsize]{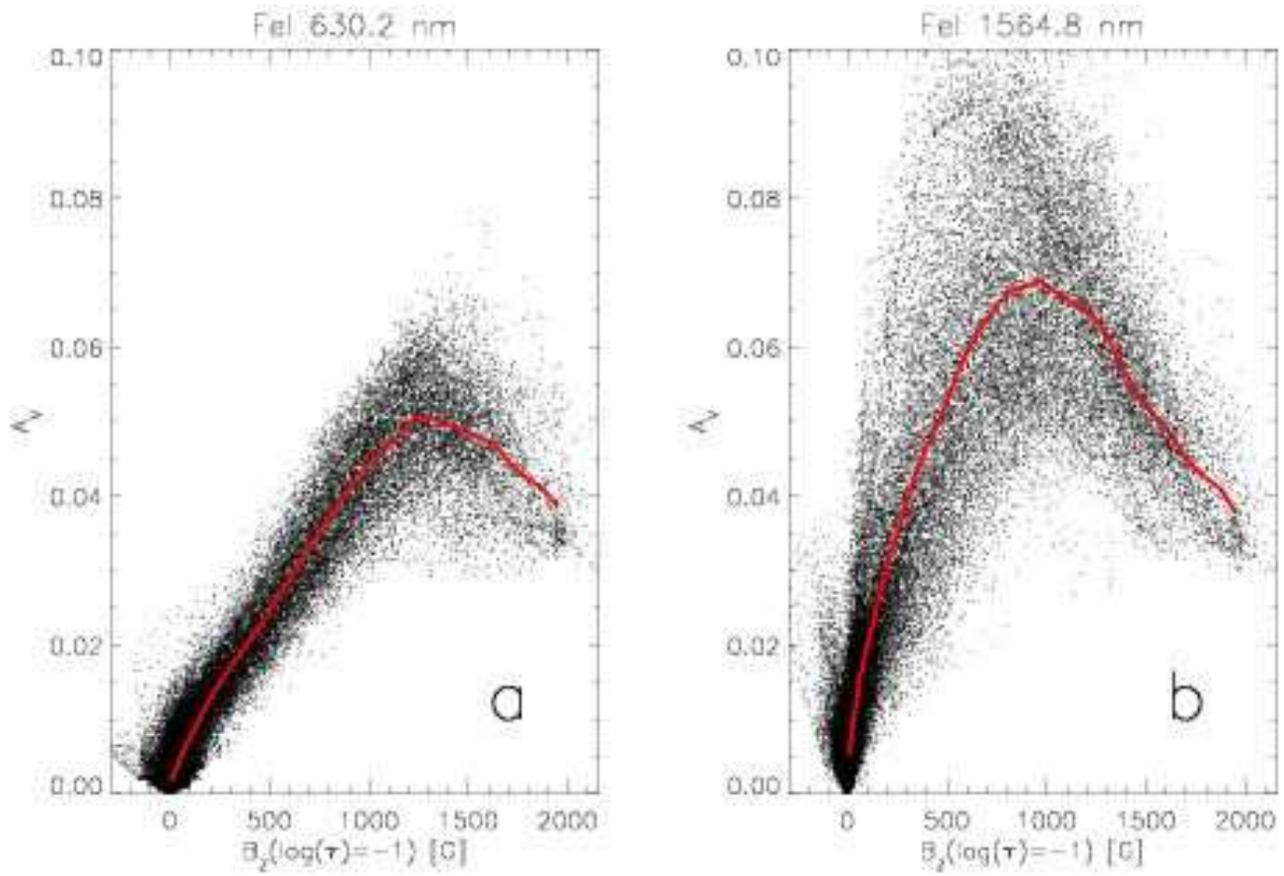}
\caption{Scatter plots of the unsigned Stokes-$V$ area versus the
vertical (line-of-sight) component of the magnetic field at
$\tau_{500}=0.1$ {\em a:} \fei\ 630.2~nm, {\em b:} \fei\ 1564.8~nm.  The
lines show binned averages.}
\label{bzstokesv}
\end{figure*}

Fig.~\ref{bzstokesv} shows scatter plots of the unsigned Stokes-$V$
area, $A_V$ versus the strength of the vertical (line-of-sight)
component of magnetic field at $\tau_{500}=0.1$. It illustrates the
different magnetic sensitivity of the two lines. The Stokes signal of
the 630.2~nm line depends linearly on the magnetic field strength until
the Zeeman splitting is no longer small compared to the line width and
saturation sets in for $B_z\ga1200~\mathrm{G}$. 
The curve for the infrared line with its higher Zeeman sensitivity
becomes nonlinear at lower field strength and bends over around
$900\,$G.
This line also exhibits a larger scatter of $A_V$ values for
intermediate field strengths. The decrease of the Stokes-$V$ area of
both lines for large field strength is due to the temperature-induced
line weakening in the strong flux concentrations
(cf. Fig.~\ref{btaustokesi}). The more rapid decrease of $A_V$ with
$B_z$ for the the 1564.8~nm line is consistent with the stronger
weakening of this line as shown in Fig.~\ref{btaustokesi}b.

\begin{figure*}
\centering
\includegraphics[width=\hsize]{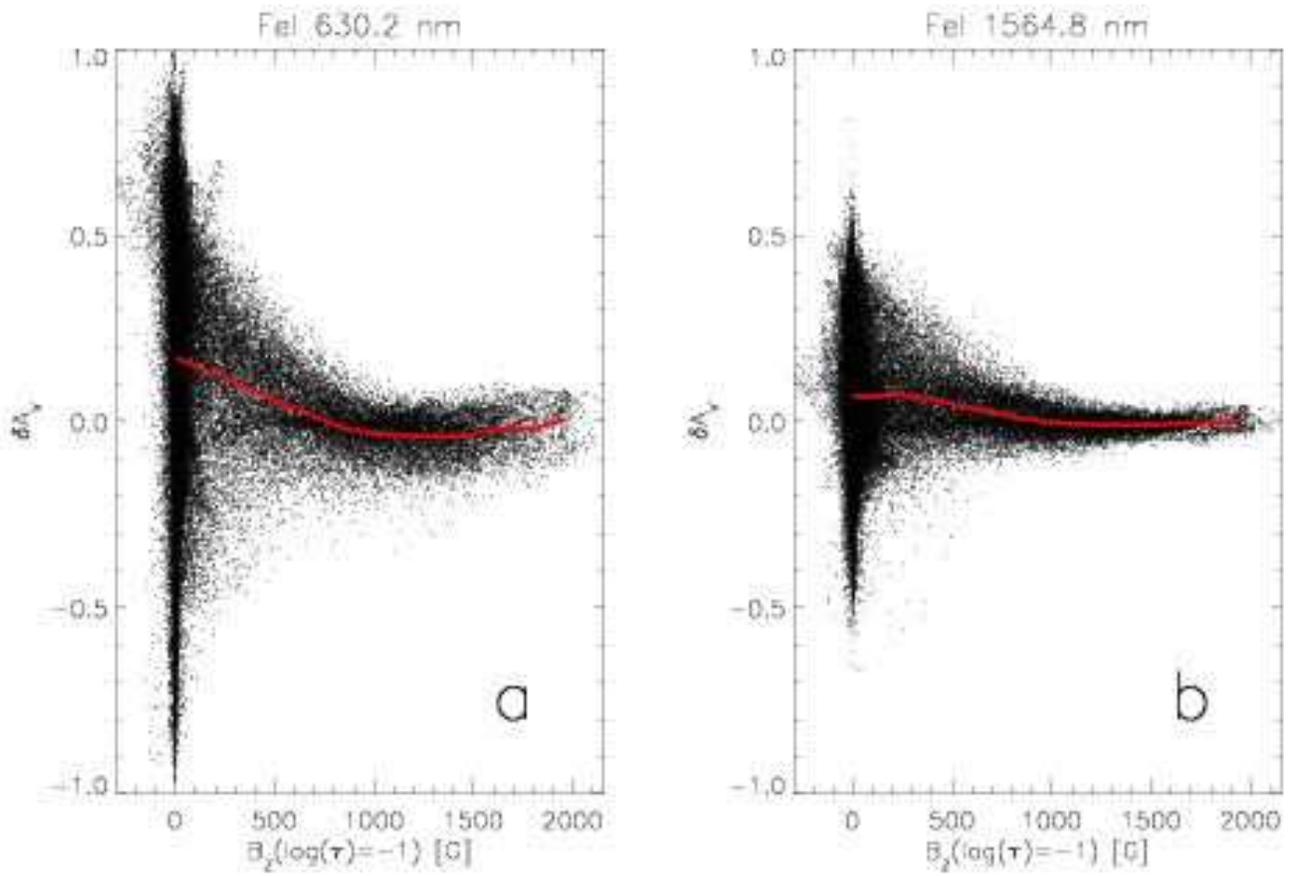}
\caption{Scatter plots of the Stokes-$V$ area asymmetry versus the
 vertical (line-of-sight) component of the magnetic field at
 $\tau_{500}=0.1$ {\em a:} \fei\ 630.2~nm, {\em b:} \fei\ 1564.8~nm.  The
 lines show binned averages.}
\label{bzvmodv}
\end{figure*}

\begin{figure*}
\centering
\includegraphics[width=\hsize]{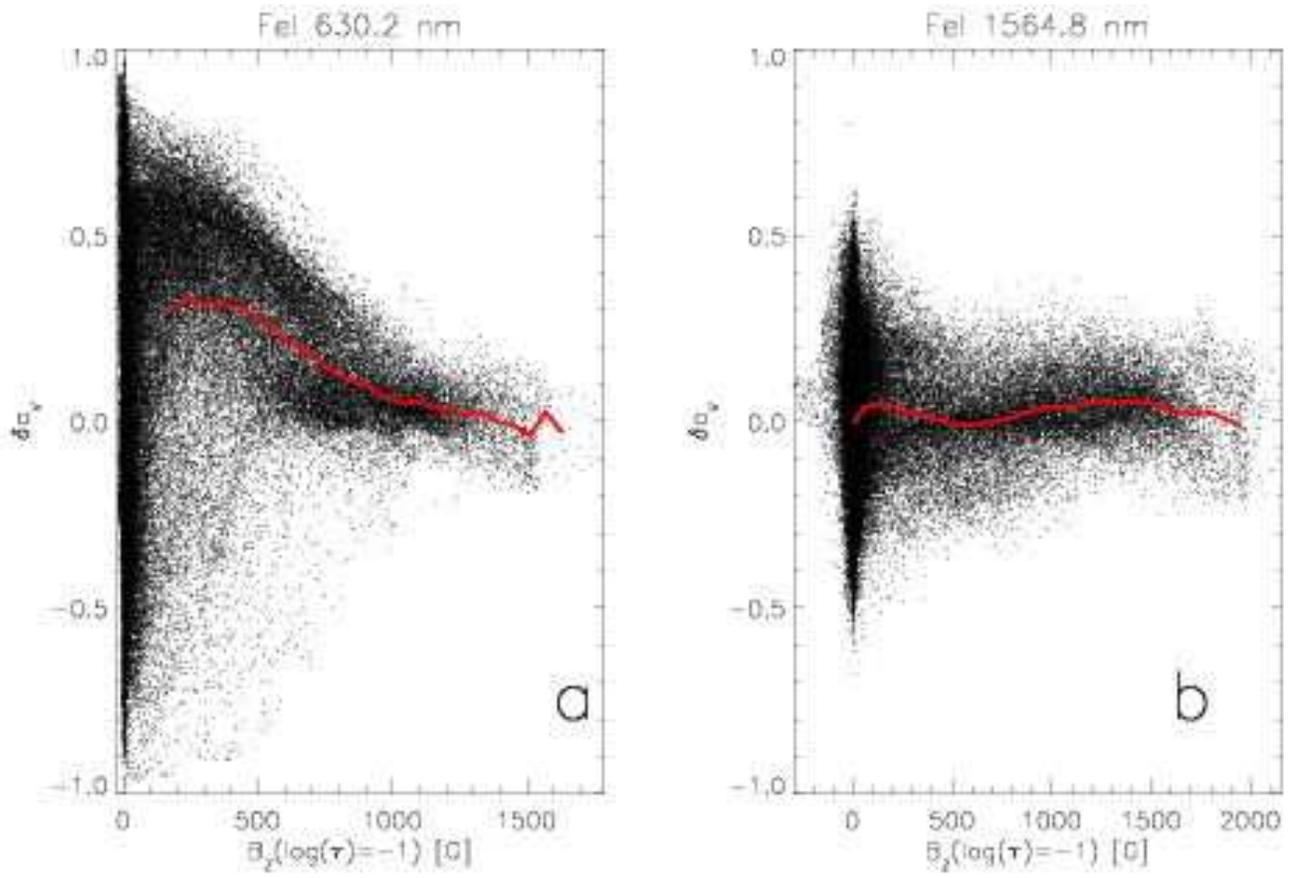}
\caption{Same as Fig.~\ref{bzvmodv} for the Stokes-$V$ amplitude asymmetry.} 
\label{ampas}
\end{figure*}

Figures~\ref{bzvmodv} and ~\ref{ampas} show scatter plots of the
Stokes-$V$ area asymmetry, $\delta A_V$, and amplitude asymetry, $\delta
a_V$, respectively, versus the vertical (line-of-sight) magnetic field
strength. The binned averages of the asymmetries given by the red curves
show that the asymmetries decrease towards stronger fields, even with a
tendency to become negative. The results for the 630.25~nm line are in
qualitative agreement with the observations of \citet[][see their
Figs.~15 and 16]{Martinez-Pillet:etal:1997} and \citet[][see their
Fig.~6]{Sigwarth:etal:1999}. Note, however, the difference in spatial
resolution between simulation and observations, so that a more
quantitative comparison is not indicated.

The asymmetry distributions shown here are in qualitative agreement
with the results of \citet{Khomenko:etal:2005b} on the basis of a MURaM
simulation of a decaying mixed-polarity field. In both cases (and in
agreement with observations) it is found that the amplitude asymmetry,
on average, is a factor 2 to 3 larger than the area asymmetry. On the
basis of calculations with an idealized flux tube model,
\citet{Bellot-Rubio:etal:1997} have suggested that a downflow in the
flux concentration, in addition to the external downflow, could lead to
asymmetries with this property.  In fact, the simulated flux
concentrations, on average, show internal downflows (see
Fig.~\ref{zcwl}), consistent with this suggestion.
 
\begin{figure*}
\centering
\includegraphics[width=\hsize]{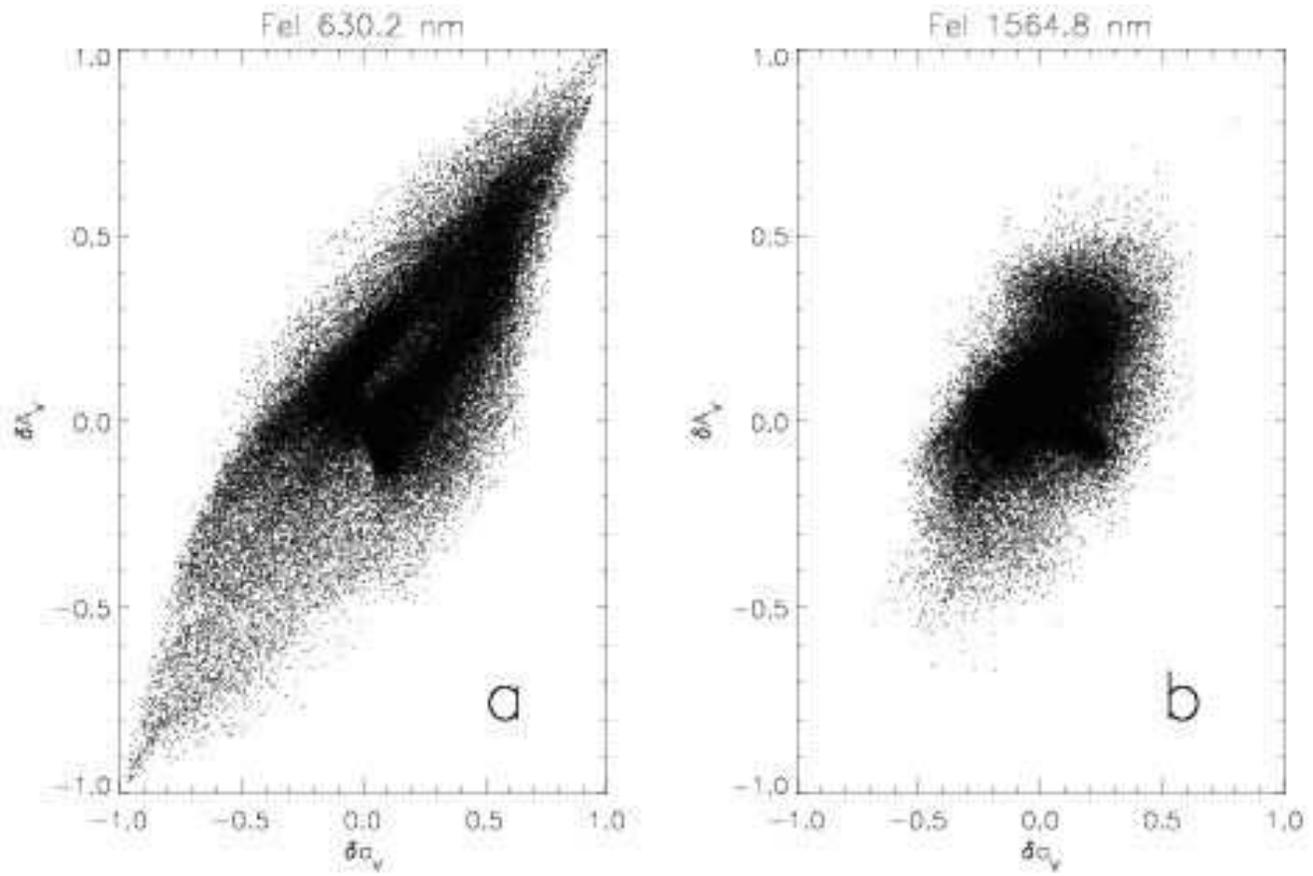}
\caption{Scatter plots of Stokes-$V$ area asymmetry versus Stokes-$V$
 amplitude asymmetry. {\em a:} \fei\ 630.2~nm, {\em b:} \fei\ 1564.8~nm.}
\label{avsa}
\end{figure*}

The asymmetries for the 1564.8~nm line are systematically smaller than
those for the visible line. 
The asymmetry is maximized if the wavelength shift due to the
Zeeman splitting is similar to the Doppler shift due to the velocity
gradient \citep{Grossmann-Doerth:etal:1988}. This is roughly the case
for the visible line while the stronger Zeeman sensitivity of the
infrared line leads to smaller asymmetry compared to that of the visible
line. Furthermore, the infrared line is weaker and less saturated than
the visible line, which additionally reduces the asymmetry
\citep{Solanki:1989}.  The large scatter of the asymmetry values for
magnetic field strengths below $\sim200\,$G is consistent with the
observed asymmetries of weak Stokes-$V$ profiles
\citep[e.g.,][]{Sigwarth:2001} and indicates that the scatter in such
observations is not entirely due to noise.
In the
simulations, many of the weak $V$ profiles result from strongly inclined
fields in the granular upflows.  Note that, according to the definitions
of the asymmetries given by Eqs.~(\ref{stvaseq}) and (\ref{stvaseqa}),
complex profiles, e.g. those with 3 or more lobes, cannot be
distinguished from normal profiles in Figs.~\ref{bzvmodv} and
\ref{ampas}. Fig.~\ref{avsa} shows that the average amplitude and area
asymmetries are well correlated, particularly for the 630.2~nm line,
which is in accordance with the observational results of \citet[][see
their Fig.~9]{Sigwarth:etal:1999}.

\begin{figure}
\includegraphics[width=\hsize]{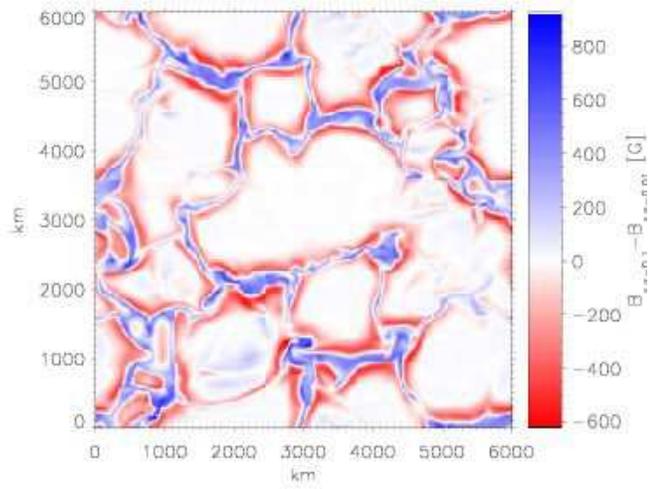} 
\caption{Spatial distribution of $\left(
B_{z,{\tau_{500}=0.1}}-B_{z,{\tau_{500}=0.01}} \right)$, representing
the average line-of-sight gradient of the vertical magnetic
field. Magnetic field increasing with optical depth is indicated by blue
color, magnetic field decreasing with optical depth is shown in red.}
\label{gradients}
\end{figure}

\begin{figure}
\includegraphics[width=\hsize]{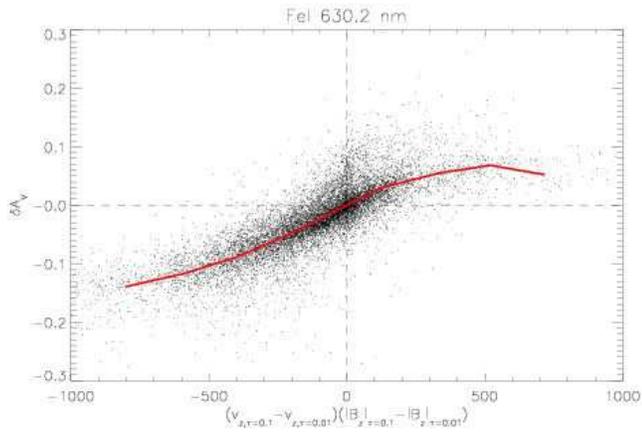}
\caption{Scatter plot of the Stokes-$V$ area asymmetry for the 630.2~nm
  line versus a quantity representing the gradients of the line-of-sight
  magnetic field and velocity within the height range of line
  formation. The sign of the bin-averaged $\delta A_V$ (indicated by the
  line) is in agreement with the relation given in Eq.~(\ref{signaseq}). }
\label{vassym}
\end{figure}

\begin{figure*}
\centering
\includegraphics[width=\hsize]{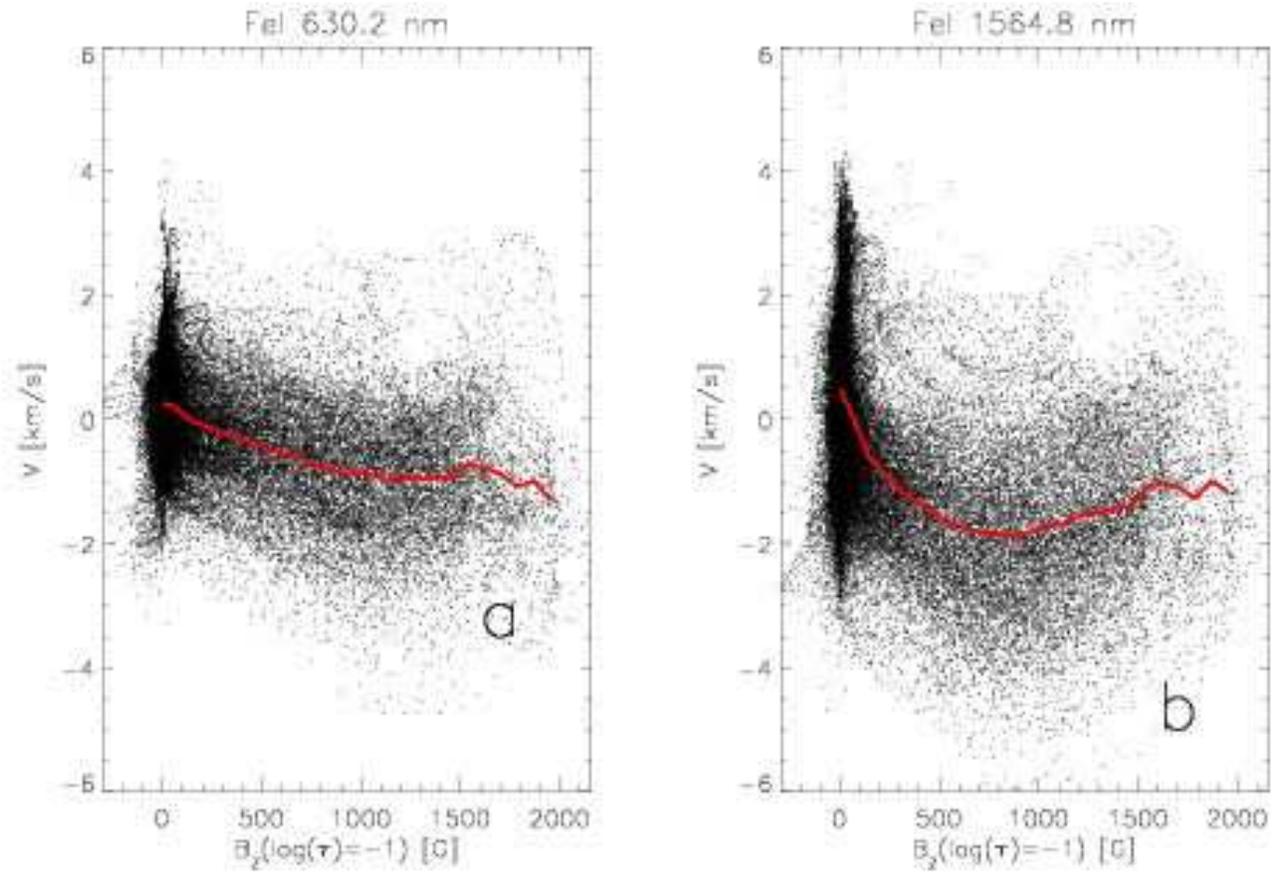}
\caption{Scatter plots of the line-of-sight velocity calculated from the
zero-crossing-wavelength of the Stokes-$V$ profile versus the vertical
component of the magnetic field at $\tau_{500}=0.1$. {\em a:} \fei\
630.2~nm, {\em b:} \fei\ 1564.8~nm.  The lines show binned
averages. On average, strong magnetic flux concentrations are associated
with downflows (negative velocity), while weak fields $B< 200\,$G
show some preference for (granular) upflows.}
\label{zcwl}
\end{figure*}

Considering the average $V$-profile of the 630.25~nm line over the whole
computational domain of 6$\,$Mm$\times6\,$Mm, we find values of
$\delta A_V=3.6$\% for the area asymmetry and $\delta a_V=17.8$\% for
the amplitude asymmetry. This can be compared with similar averages
(albeit with different averaging areas) of observed plage spectra by
\citet[][see their Fig.~14]{Martinez-Pillet:etal:1997}, who find $\delta
A_V=3.5...4.5$\% and $\delta a_V=9...11$\%. and by \citet[][see their
Fig.~13]{Sigwarth:etal:1999}, who give values of $\delta A_V=3$\% and
$\delta a_V=11$\%. The dominance of positive asymmetries (i.e., stronger
blue wing of the Stokes-$V$ profile) for intermediate field strengths in
Figs.~\ref{bzvmodv} and ~\ref{ampas} supports the interpretation that
these average asymmetries originate in the peripheral parts of magnetic
flux concentrations, which expand with height and are surrounded by
strong downflows (such as the example shown in Figs.~\ref{flt2dcut} and
\ref{088200prof}; see also Table~1). Lines of sight in the periphery of
a flux concentration cut through the magnetopause and thus exhibit a
reversal of gradient of the line-of-sight magnetic field (field strength
decreasing with depth) in comparison to the central parts of the flux
concentrations, where the field strength increases with depth (see
Fig.~\ref{gradients}). Together with the increase of the (downflow)
velocity with depth, this leads to positive asymmetries of the
Stokes-$V$ profiles originating in the peripheral parts of the flux
concentrations (see the central frame of
Fig.~\ref{images088200}. According to Eq.~(\ref{signaseq}), the sign of
the area asymmetry should depend on the product of the gradients of
magnetic field and velocity along the line of sight. Fig.~\ref{vassym}
shows the dependence of the Stokes-$V$ area asymmetry for the 603.2~nm
line on the product $\left( v_{z,\tau_{500}=0.1}-v_{z,\tau_{500}=0.01}
\right) \cdot \left( |B_z|_{\tau_{500}=0.1}-|B_z|_{\tau_{500}=0.01}
\right)$, which corresponds to the average gradients over the formation
height of the lines, thus representing the relevant quantity in
Eq.~(\ref{signaseq}). It is clear that the sign of the average asymmetry
is in accordance with the prediction from this relation. The scatter
results from the fact that the quantity considered in Fig.~\ref{vassym}
represents average gradients over an extended height range and thus does
not take into account local variations within the height range of line
formation. The numerous points with negative area asymmetry in
Fig.~\ref{vassym} in most cases correspond to very weak Stokes-$V$
profiles formed over granules; they give only a minor contribution to
the asymmetry of the average profile.

The wavelength shift of the Stokes-$V$ zero crossing has often been used
to determine the velocity of magnetized plasma in observations
\citep[e.g.,][]{Solanki:1986, Grossmann-Doerth:etal:1996,
Martinez-Pillet:etal:1997, Sigwarth:etal:1999}. The 630.2~nm and
1564.8~nm lines are formed at somewhat different heights in the solar
atmosphere, so that they trace the velocity field along the line of
sight in a different way.  The 1564.8~nm line is formed deeper, where
velocities typically are larger.  For the simulated magneto-convection,
Fig. \ref{zcwl} shows scatter plots of the line-of-sight velocity
determined from the zero-crossing wavelength of the synthetic Stokes-$V$
profiles versus the vertical magnetic field at the level
$\tau_{500}=0.1$. The binned averages given by the line show that
downflows (negative velocity) dominate for field strength above a few
hundred Gauss, with the stronger flows typically shown by the infrared
line. For field strength around one kilogauss we find average downflows
of the order of 1~km$\,$s$^{-1}$ with the 630.2~nm line
\citep[cf.][their Fig.~6]{Sigwarth:etal:1999} while the deeper
originating 1564.8~nm shows about 2~km$\,$s$^{-1}$ downflow. The weak
fields are predominantly associated with granular upflows.  The
strongest redshifts shown by the infrared line are at field strengths in
the range 500--1000~G, which are typical of the expanding field in the
periphery of strong flux concentrations. For field strengths
corresponding to the cores of the flux concentrations, the zero-crossing
shift is reduced to about 1~km$\,$s$^{-1}$. This suggests that this
shift could be partly due to the entrainment of material from the
surrounding granular downflow, an effect probably depending also on the
spatial resolution of the simulation.

\section{Conclusions}

We have analyzed the spectro-polarimetric signatures of 3D MHD models of
solar magneto-convection, based on the synthesis of Zeeman-sensitive
\fei\ lines in the visible and in the infrared. The comparison between
various properties of the synthetic Stokes profiles (such as line
strength, line width, Doppler shift, Stokes-$V$ signal, area and
amplitude asymmetries) with the physical properties of the model
(temperature, magnetic field, and velocities) provides insight into the
mechanisms underlying various characteristics of the observed
spectra. In particular, the generation of strongly asymmetric Stokes-$V$
profiles in the periphery of magnetic flux concentrations largely
confirms earlier simplified models based upon lines of sight crossing
the magnetopause of flux concentrations fanning out with height.
The line weakening in strong magnetic fields is found to be related to
larger temperature (and smaller temperature gradient) in the simulated
magnetic flux concentrations, in agreement with earlier interpretations
of observational results.

\acknowledgements{The authors are grateful to 
Sanja Danilovic for test calculations of spectral line profiles.}


\bibliography{pap_rev_s.bbl}

\clearpage

\clearpage

\clearpage

\clearpage

\clearpage

\clearpage

\clearpage

\clearpage

\clearpage

\clearpage

\clearpage

\clearpage

\clearpage

\clearpage

\clearpage

\clearpage

\clearpage

\clearpage

\clearpage

\end{document}